\documentclass[a4paper,11pt]{article}
\pdfoutput=1 

\usepackage{jcappub} 

\usepackage[T1]{fontenc} 

\usepackage{xcolor, soul}

\usepackage{upgreek}
\usepackage{booktabs}

\usepackage{multirow}
\usepackage{subcaption}
\usepackage{siunitx} 
\title{\boldmath  Improving ANAIS--112 sensitivity to DAMA/LIBRA signal with machine learning techniques}

\author[a,b]{I.~Coarasa,}
\author[a]{J.~Apilluelo,}
\author[a,b]{J.~Amar{\'e},}
\author[a,b]{S.~Cebri{\'an},}
\author[a,b]{D.~Cintas,}
\author[a,b]{E.~Garc\'{\i}a,}
\author[a,b,c]{M.~Mart\'{\i}nez}
\author[a,b,d]{M.A. Oliv{\'a}n,}
\author[a,b,e]{Y.~Ortigoza,}
\author[a,b]{A.~Ortiz~de~Sol{\'o}rzano,}
\author[a,b]{T.~Pardo,}
\author[a,b]{J.~Puimed{\'o}n,}
\author[a,b]{A.~Salinas,}
\author[a,b]{M.L.~Sarsa,}
\author[a]{and P.~Villar}

\affiliation[a]{Centro de Astropart\'{\i}culas y F\'{\i}sica de Altas Energ\'{\i}as (CAPA), Universidad de Zaragoza, Pedro Cerbuna 12, 50009 Zaragoza, Spain}
\affiliation[b]{Laboratorio Subterr\'aneo de Canfranc, Paseo de los Ayerbe s.n., 22880 Canfranc Estaci\'on, Huesca, Spain}
\affiliation[c]{Fundaci\'on ARAID, Av. de Ranillas 1D, 50018 Zaragoza, Spain}
\affiliation[d]{Fundaci\'on CIRCE, Av. de Ranillas 3D, 50018 Zaragoza, Spain}
\affiliation[e]{Escuela Universitaria Polit\'ecnica de La Almunia de Do\~{n}a Godina (EUPLA), Universidad de Zaragoza, Calle Mayor 5, La Almunia de Do\~{n}a Godina, 50100 Zaragoza, Spain}

\emailAdd{icoarasa@unizar.es}
\emailAdd{japilluelo@unizar.es}
\emailAdd{amare@unizar.es}
\emailAdd{scebrian@unizar.es}
\emailAdd{cintas@unizar.es}
\emailAdd{edgarcia@unizar.es}
\emailAdd{mariam@unizar.es}
\emailAdd{maolivan@unizar.es}
\emailAdd{ortigoza@unizar.es}
\emailAdd{alfortiz@unizar.es}
\emailAdd{tpardo@unizar.es}
\emailAdd{puimedon@unizar.es}
\emailAdd{salinas@unizar.es}
\emailAdd{mlsarsa@unizar.es}
\emailAdd{pvillar@unizar.es}

\newcommand{\DL}{DAMA\slash LIBRA\ }
\newcommand{\ANAIS}{ANAIS--112\ }
\newcommand{\COSINE}{COSINE--100\ }

\newcommand{\ckkd}{c/keV/kg/d}

\newcommand{\Na}{$^{22}$Na\ }
\newcommand{\K}{$^{40}$K\ }
\newcommand{\NaK}{$^{22}$Na\slash$^{40}$K\ }
\newcommand{\Cd}{$^{109}$Cd\ }
\newcommand{\Cf}{$^{252}$Cf\ }

\newcommand{\Pb}{$^{210}$Pb\ }

\newcommand{\Sen}{\mathcal{S}}

\abstract{
The \DL observation of an annual modulation in the detection rate compatible with that expected for dark matter particles from the galactic halo has accumulated evidence for more than twenty years. It is the only hint of a direct detection of the elusive dark matter, but it is in strong tension with the negative results of other very sensitive experiments, requiring {\it ad-hoc} scenarios to reconcile all the present experimental results. Testing the \DL result using the same target material, NaI(Tl), removes the dependence on the particle and halo models and is the goal of the \ANAIS experiment, taking data at the Canfranc Underground Laboratory in Spain since August 2017 with 112.5~kg of NaI(Tl). At very low energies, the detection rate is dominated by non-bulk scintillation events and careful event selection is mandatory. This article summarizes the efforts devoted to better characterize and filter this contribution in \ANAIS data using a boosted decision tree (BDT), trained for this goal with high efficiency. We report on the selection of the training populations, the procedure to determine the optimal cut on the BDT parameter, the estimate of the efficiencies for the selection of bulk scintillation in the region of interest (ROI), and the evaluation of the performance of this analysis with respect to the previous filtering. The improvement achieved in background rejection in the ROI, but moreover, the increase in detection efficiency, push the \ANAIS sensitivity to test the \DL annual modulation result around 3$\sigma$ with three-year exposure, being possible to reach 5$\sigma$ by extending the data taking for a few more years than the scheduled 5 years which were due in August 2022.}

\begin{document}
\maketitle
\flushbottom

\section{Introduction}
\label{sec:intro}

For almost a century, evidence on the existence of an unknown matter component in the Universe, from the scale of galaxies and clusters of galaxies up to cosmological scales, has been steadily accumulating. The $\Lambda$CDM cosmological model is able to satisfactorily explain all the observations at the cost of introducing 84\% of the matter in the form of Dark Matter (DM)~\cite{Workman:2022ynf,Bertone:2016nfn}. Moreover, 68\% of the total matter-energy should be in the form of Dark Energy (DE). Understanding the nature of the DM has shown to be a challenge, remaining one of the most puzzling unresolved questions in Fundamental Physics. Progressing in this quest requires following complementary approaches from the different points of view of cosmology, astronomy, astrophysics, and nuclear and particle physics, and both from theory and experiment. Despite the large efforts invested in the searches for new particles with the right properties to be this DM, by combining searches at accelerators~\cite{Buchmueller:2017qhf}, indirect~\cite{Conrad:2017pms} and direct searches~\cite{Liu:2017drf,Billard:2021uyg}, the solution to the DM puzzle remains beyond our present sensitivities. 

DM direct detection experiments aim to identify the small energy deposited in a convenient target material by the interaction of the DM particles filling the Milky Way's halo. The expected signal is strongly dependent on the particle DM model and galactic halo model\footnote{In particular, the distribution of velocities of the DM particles and the particle density at the Solar System position.}, being either unknown or affected by large uncertainties. In the preferred DM scenarios, these interactions should produce nuclear recoils by the elastic scattering off the target nuclei of the detector. In spite of the wide experimental effort, after more than three decades of DM direct searches, most of the experimental results are compatible with the estimated backgrounds~\cite{Billard:2021uyg,Schumann2019}. Only the \DL experiment, which uses NaI(Tl) crystal detectors at the Gran Sasso National Laboratory (LNGS) in Italy, has provided a long-standing positive result~\cite{DAMA:2000mdu, DAMA:2008jlt,DAMA:2010gpn,Bernabei:2013xsa,Bernabei:2018jrt,Bernabei:2020mon,Bernabei:2021kdo}, the observation of a highly statistically significant annual modulation in the low energy detection rate (below 6~keV$_{\textnormal{ee}}$\footnote{Electron-equivalent energy. In the following, we will use just keV for keV$_{\textnormal{ee}}$.}), which is compatible with that expected for DM particles distributed following the standard halo model~\cite{Drukier:1986tm,Freese:1987wu}. The \DL result has not been reproduced by any other experiment and, for years, the most plausible DM scenarios have been strongly constrained by the results of other experiments using different target materials and detection techniques~\cite{XENON:2017nik,XENON:2018voc,XENON:2019gfn,LUX:2016ggv,LUX-ZEPLIN:2022qhg,PandaX-II:2017hlx,PandaX-4T:2021bab,XMASS:2018xyo,DEAP:2019yzn,DarkSide:2018bpj,DarkSide:2018kuk,SuperCDMS:2014cds,SuperCDMS:2018gro,EDELWEISS:2019vjv,CRESST:2019jnq,CRESST:2022jig,kims2014,PICO:2019vsc,CDEX:2019aqn}. A model-independent test of the \DL result requires an experiment using the same target material, a similar (or better) background and threshold energy, large exposure (product of mass and measurement time) and stability, in order to analyze the annual modulation effect. This is the goal of the ANAIS (Annual modulation with NaI Scintillators) experiment~\cite{anais2019Perf,Amare:2022ncr}: to provide a model independent confirmation or refutation of the annual modulation positive signal reported by \DL using the same target and technique, but different experimental conditions. \ANAIS started taking data at the Canfranc Underground Laboratory (LSC), in Spain, in August 2017 with a mass of 112.5~kg and the last reported annual modulation results, corresponding to 3 years of data taking, are not compatible with DAMA/LIBRA, for a sensitivity of 2.5-2.7$\sigma$, depending on the energy region~\cite{Amare:2021yyu}. 

Other projects worldwide share the ANAIS' goal. The \COSINE experiment is taking data at YangYang laboratory, in South Korea since September 2016 with an effective mass of $\approx$~60~kg. \COSINE has released model-dependent results excluding DAMA/LIBRA--compatible parameters regions~\cite{COSINE-100:2021xqn}, while the last annual modulation analysis reported point to a compatibility with both DAMA/LIBRA and absence of modulation~\cite{COSINE-100:2021zqh}. Other projects in R\&D phase are SABRE~\cite{SABRE:2018lfp,Calaprice:2021yml}, PICOLON~\cite{picolon2021} and COSINE--200, aiming at improving the radiopurity of the NaI(Tl) crystals using an only-scintillation detection approach, whereas the COSINUS project~\cite{cosinus2016} bases the detection on the simultaneous measurement of the light and phonons, which allows identifying the interacting particle by the different sharing between the two energy conversion channels. Previously, DM-Ice17, operating at the South Pole, tried to address this goal, but its sensitivity was too low~\cite{DM-Ice:2016snk}.
 
ANAIS sensitivity is limited by anomalous scintillation events which dominate the detection rate below 10~keV. Robust protocols for the selection of events associated with the NaI(Tl) scintillation in the crystal bulk, as DM particles should produce, are then mandatory. In this paper, a complete description of a new machine learning procedure developed for the selection of low-energy bulk scintillation events in \ANAIS data is presented. The previously used analysis protocol showed a very low acceptance efficiency at energies below 2~keV and, in addition, resulted in a background excess mostly below 2~keV which could be attributed to non-scintillation events leaking the filtering procedure~\cite{anais2019Perf}. This motivated the search for more powerful filtering techniques to be applied to \ANAIS data, with the consequent improvement in sensitivity. 

Machine learning techniques applied in other experiments searching for rare events have demonstrated to be very powerful in discriminating signal from noise~\cite{edelweiss2018,xenon1t2020,LUX:2020yym,COSINE-100:2020wrv,DRIFT:2021uus,Herrero-Garcia:2021goa}. Following such a different strategy for the filtering of \ANAIS data we expected both, to improve acceptance efficiencies for bulk scintillation events (signal) and better rejection of non-bulk scintillation events (noise) between 1 and 2~keV. For that goal, we have implemented a supervised machine learning technique based on a multivariate analysis, in particular a boosted decision tree (BDT). In a previous work~\cite{ivanThesis,anais2021ML}, a two-phase training was performed: \Cd calibration events were considered signal events while noise samples were built from a blank module (i.e., identical to the \ANAIS ones, but without scintillating NaI(Tl) crystal) events in the first phase, see Section~\ref{ssec:blank}, and background events in the second phase. In this paper, we report on the update of this BDT approach, profiting from the development of a dedicated neutron calibration program for the \ANAIS experiment, see Section~\ref{ssec:neucal}. We will use as signal events for the algorithm training scintillation events in the bulk of the crystals produced by neutron interactions, which are mostly associated to elastic nuclear recoils in that energy region, and as noise events, those coming from the blank module, without requiring to use background events at all.

The structure of the article is the following: some relevant features of the \ANAIS experimental set-up, in particular the complementary blank module set-up and the dedicated neutron calibration program developed in the last year of operation, are described in Section~\ref{sec:anais}; the machine learning algorithm, training populations and parameters selected and the algorithm response are presented in Section~\ref{sec:ML}; Section~\ref{sec:evsel} describes the event selection based on the machine learning algorithm output; in Section~\ref{sec:valid}, the validation of the machine learning response is checked;  the sensitivity improvement to the annual modulation observed by \DL is covered in Section~\ref{sec:sens}. Finally, conclusions are summarized in Section~\ref{sec:conclusions}.


\section{The \ANAIS experiment}
\label{sec:anais}


\subsection{The apparatus}

\ANAIS consists of nine 12.5~kg NaI(Tl) modules in a 3$\times$3 matrix configuration, for a total mass of 112.5~kg. The ANAIS modules were manufactured by Alpha Spectra Inc. in Colorado (US) between 2012 and 2017. All the crystals are cylindrical (4.75'' diameter and 11.75'' length) and are housed in OFE (Oxygen Free Electronic) copper; the encapsulation has a Mylar window in the lateral face for low energy calibration using external gamma sources. Every crystal was coupled through synthetic quartz windows to two highly efficient Hamamatsu R12669SEL2 photomultipliers (PMTs) at the LSC clean room. The high optical quality of the crystals added to the high-efficiency PMTs result in remarkable light collection at the level of 15~photoelectrons/keV~\cite{anais2019Perf,anais2017LY} in all the modules.

\begin{figure}
    \centering
    \includegraphics[width=0.50\textwidth]{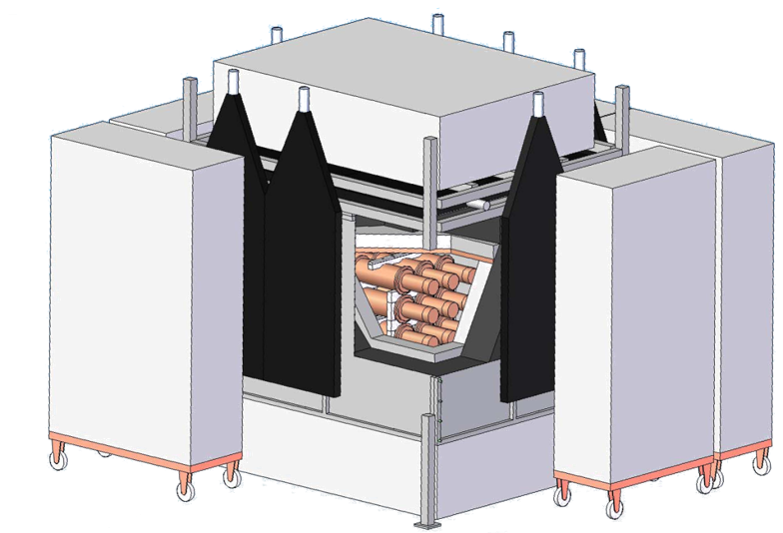}
    \hspace{0.05\textwidth}
    \includegraphics[width=0.40\textwidth]{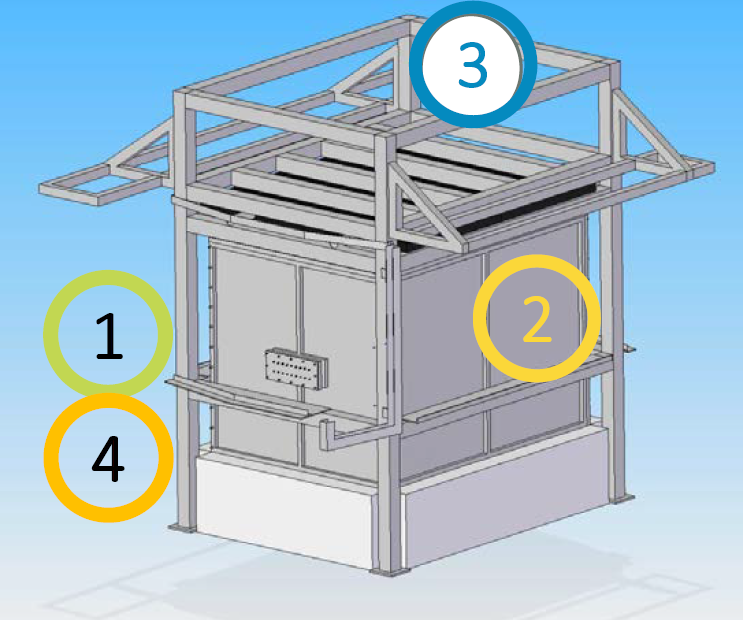}
    \caption{Left: artistic view of the \ANAIS set-up at LSC showing the different shielding components (see text for more details). Right: position of the $^{252}$Cf neutron source during the different calibration runs carried out: west side  (1 and 4), south side (2) and top side (3). The detector axis is aligned parallel to the east-west direction. The source is placed outside the anti-radon box and lead shielding, but inside the muon vetoes and neutron shielding.}
    \label{fig:neucal1}
\end{figure}

The \ANAIS shielding consists of 10~cm of archaeological lead, 20~cm of low activity lead, an anti-radon box (continuously flushed with radon-free nitrogen gas), an active muon veto system made up of 16~plastic scintillators covering the top and sides of the whole set-up and 40~cm of neutron moderator (a combination of water tanks and polyethylene blocks). The left panel of Figure~\ref{fig:neucal1} shows a design of the set-up. The hut housing the experiment is located at the hall B of LSC under a rock overburden equivalent to 2450~m of water. A more detailed description of the \ANAIS experimental set-up can be found in \cite{anais2019Perf}.
%

Concerning the \ANAIS data acquisition system, each PMT signal is fully processed by the electronic chain and divided into a trigger signal, a low-energy (LE) signal and a high-energy (HE) signal. Triggering of every detector is done by the coincidence (logical AND) of the two PMT signals of any detector at photoelectron (phe) level in a 200~ns window, whereas the main acquisition trigger is the logical OR of individual detectors. The HE signals are fed into CAEN V792 QDC (charge-to-digital-converter) modules, with 1~$\upmu$s integration window and 12 bits resolution. The LE signals of each PMT are sent to MATACQ-CAEN V1729A digitizers, which sample the waveforms at 2~GS/s in a 1.2~$\upmu$s window with high resolution (14 bits). In the analysis phase, a software trigger position ($t_0$) is determined for every waveform by using a threshold level over the baseline, independently from the hardware trigger threshold. $t_0$ is used later in the calculation of several pulse shape parameters (see Sections~\ref{ssec:filtering} and \ref{sec:ML}).


\subsection{The analysis}
\label{ssec:filtering}

The main analysis parameters are drawn from the waveform of LE pulses. In particular, the pulse area (our energy estimator) is calculated by integrating the pulse from the software trigger position, individually for each PMT. Moreover, peaks in the waveform of each PMT are identified using a peak-searching algorithm~\cite{maThesis} based on a low pass filter and the detection of a sign change in the derivative waveform. The number of peaks in the waveform of every photomultiplier ($n_0$ and $n_1$ for PMT0 and PMT1, respectively) are stored, besides the corresponding positions in the waveform. These variables, as it will be commented below, are very relevant in the \ANAIS analysis. Other analysis parameters, in particular those related with the pulse shape, will be introduced in following sections. 

The response of the \ANAIS modules is monitored through the PMTs gain and the light collection. In particular, the PMT gain is directly monitored by the charge distribution of a single photoelectron for each PMT, the so-called single electron response (SER). We estimate periodically, every two weeks approximately, this SER for each PMT by selecting the last peak identified in the tail of PMT pulses having a very low number of peaks. 

At very low energy, the number of peaks in the pulse corresponds to the number of photoelectrons. However, as energy increases, overlapping of photoelectron signals in time produces a reduction of the number of identified peaks. Because of this, we define the number of photoelectrons ($nphe$), independently from the number of peaks, as the pulse area divided by the mean SER area for each PMT ($nphe_0$ and $nphe_1$ for PMT0 and PMT1 signals, respectively), and then we add both to build the total number of photoelectrons for each module, either the blank module or the \ANAIS modules. Some examples of pulses with a low number of phe can be found in Section~\ref{ssec:blank}, Figure~\ref{fig:blankpulses}.

The stability of the \ANAIS experiment is monitored by calibrating with external \Cd sources every two weeks. All the nine modules are simultaneously calibrated using a multi-source system built in a flexible wire, which allows the sources to be introduced from the outside, keeping the system closed and guaranteeing radon-free operation. The energy calibration of the region of interest (ROI, 1-6~keV) is carried out by combining the information from those \Cd lines and from known lines present in the background at 3.20 and 0.87~keV (following the electron capture decay of \K and \Na from bulk crystal contamination, respectively). The events corresponding to those background lines are tagged by coincidences with high energy depositions (1460.8 and 1274.5~keV, respectively) in a second module. The use of \NaK lines increases the reliability of the energy calibration in the ROI since they are actually either in the ROI or very close to the analysis energy threshold, set at 1~keV~\cite{anais2019Perf}. Moreover, \NaK events provide a clean sample of very low energy depositions converted into scintillation in the crystal bulk unlike \Cd events, which because of the external origin of the gammas, correspond to scintillation produced near to the surface in the middle position between both PMTs. This makes this \NaK population very interesting in the \ANAIS analysis.

The \ANAIS background in the ROI is dominated by non-bulk scintillation events~\cite{anais2019Perf}, requiring the development of robust filtering protocols (based on the pulse shape and light sharing among the two PMTs) to reduce it. The fine tuning of these event rejection procedures was performed unblinding the $\sim$10\% of the data from the first year of measurement of the experiment, chosen randomly distributed days along the data taking period. The efficiency of these selection criteria was calculated with populations of scintillation events (from $^{109}$Cd, $^{40}$K, and $^{22}$Na) in \cite{anais2019Perf}, showing to be very close to 100\% down to 2~keV, but then decreasing steeply to about 15\% at 1~keV. However, the \ANAIS modules are effectively triggering below 1~keV, as \NaK coincident events show~\cite{anais2019Perf}.  

The origin of these non-bulk scintillation events is not fully established, but undoubtedly they are associated to the detection of light in the PMTs and not to spurious coincidences between individual photoelectrons with origin in the PMTs dark current. Some of those events, typically fast and asymmetric in the light sharing, are originated by Cherenkov emission in the glass structure of the PMTs. Nevertheless, other kind of light emissions are observed, which could be produced at the dynodes of the PMTs, quartz windows, optical coupling media, Teflon wrapping, etc. 

The filtering protocols initially developed and applied in \ANAIS have been exhaustively described in \cite{anais2019Perf} and they are based on standard cuts on a few parameters. To be accepted, an event must fulfill the following criteria:
\begin{enumerate}
    \item pulse shape according to NaI scintillation time constant. A bi-parametric pulse shape cut, combining the fraction of the pulse area in [100, 600]~ns after the event trigger ($P_1$, see Equation~\ref{eq:P1}) and the logarithm of the amplitude-weighted mean of the arrival times corresponding to the identified peaks in the pulse in the digitized window ($\mu_p$, see Equation~\ref{eq:mup}) removes fast Cherenkov-like events originated in the PMTs, random coincidences between individual dark current photoelectrons and long phosphorescences in the crystal or pulse tails.
    \item symmetrical sharing of light between the two PMTs. More than 4~peaks in each PMT signal ($n_0$>4 and $n_1$>4) are required, removing events having a strongly asymmetric light sharing, with a very low number of peaks in one of the PMTs. These asymmetric events do not have the timing characteristics of afterpulses or pulse tails, and are observed below 2~keV.
    \item single-hit events. Any event showing coincident energy depositions in more than one \ANAIS module is produced by the residual cosmic ray muon flux, radioactive contaminants in the set-up components or environmental radioactivity, but not by DM particles.
    \item events arriving more than 1~s after the last muon veto trigger. As \ANAIS trigger is set at the photoelectron level, photons in the tail of pulses caused by the passage of a very energetic particle are able to trigger the acquisition multiple times, up to hundreds of ms after the original pulse onset. At the same time, other energy depositions associated to delayed processes following the muon interactions in the shielding could be rejected. Removing muon related events is very important for the \ANAIS goal, because the muon flux in deep underground laboratories has shown annual modulation~\cite{Borexino:2018pev,OPERA:2018jif}.
\end{enumerate}

Although this filtering procedure works very well above 2~keV, in the region from 1 to 2~keV it shows some weaknesses. On the one hand, as commented above, the efficiency of acceptance of bulk scintillation events decreases very sharply, reducing the sensitivity, and on the other hand, the measured rate in this region is about 50\% higher than expected according to the \ANAIS background model~\cite{anais2019Bkg}, pointing at a possible leaking of non-bulk scintillation events as responsible of that excess.


\subsection{The neutron calibration program}
\label{ssec:neucal}

As stressed in Section~\ref{sec:intro}, a comparison between results obtained with different target nuclei in DM searches is affected by model-dependencies. Using the same target should remove these dependencies completely. For instance, the expected contribution of energy released by Migdal-electrons should be exactly the same for detectors consisting of the same nuclei~\cite{Ibe:2017yqa}. There is, however, one caveat: a direct comparison between experiments using the same target is direct only in the case in which the full response function of the detectors is taken into account in such a comparison of results. This is particularly relevant in the testing of the DAMA/LIBRA result because scintillation is strongly quenched for energy deposited by nuclear recoils with respect to the same energy deposited by electrons, and experiments are usually calibrated (i.e., the response function is calculated) using gamma/electron radiation. Because of this, experimental measurements are often presented in terms of electron-equivalent energy (keV$_{\textnormal{ee}}$) and experiments can be directly compared only if energy is deposited by this channel. In the case when DM particles produce nuclear recoils, quenching factors (QF) should be well known to re-calibrate the energy depositions scale.

This is not the case for NaI(Tl) scintillators. Measurements of QF in the energy region relevant for DM searches, and in particular for the comparison with DAMA/LIBRA in case the signal is attributed to nuclear recoils, do not agree, and in fact are affected by strong discrepancies (QF dependent on the energy versus constant values, for instance) \cite{Cintas:2021fvd,Bignell:2021bjx,Joo:2017aws,Stiegler:2017kjw,Xu:2015wha,Collar:2013gu,Chagani:2008in,Gerbier:1998dm,Tovey:1998ex,Bernabei:1996vj,Spooner:1994ca}.

The \ANAIS experiment's design included periodical calibration using \Cd sources to monitor and correct any gain drift that could compromise the experimental goals. Stability in the long term of any experiment searching for an annual modulation in the data at very low energies is mandatory. Such \Cd calibrations are carried out every two weeks and produce lines at 11.9, 22.6 and 88.0 keV, the first of them attributed to Br-content in the source housing. Besides gain monitoring purposes, these calibrations provided a clean population of low energy events corresponding to scintillation in the NaI(Tl) crystal which were used to determine the efficiency of the filtering applied for the 1.5, 2 and 3 years analyses~\cite{Amare:2019jul,Amare:2019ncj,Amare:2021yyu}. 

However, in order to fully understand the response of the \ANAIS detectors to different particle energy depositions, a specific neutron calibration program was necessary. This program, under development, combines two different approaches: on the one hand, sodium quenching factor measurements were carried out at TUNL (Duke University, North Carolina, US) using small crystals from the same growing batches of the \ANAIS crystals  (see \cite{Cintas:2021fvd} for preliminary results); on the other hand, neutron calibrations {\it onsite} have been performed, starting in 2021, using \Cf sources of low activity\footnote{The \Cf source used for the calibrations referred in the text had an activity of 3~kBq on April 2021.} at LSC. This paper focuses on the latter, which provide a clean population of bulk scintillation events, dominated by nuclear elastic scattering, homogeneously distributed in all the crystals volume and that can be further clean-up by selecting only multiple-hit events. Because the low-energy dependence of the elastic scattering differential spectrum is exponential-like, a large number of events in the ROI is obtained in only a few calibration runs, lasting 3--4~h each. Up to the moment of writing this article, four calibration runs have been carried out, distributed along one year and a half and in different positions of the set-up (see Figure~\ref{fig:neucal1}, right panel).

\begin{figure}
  	\begin{subfigure}[b]{0.5\textwidth}
      \centering
  	  \includegraphics[width=\textwidth]{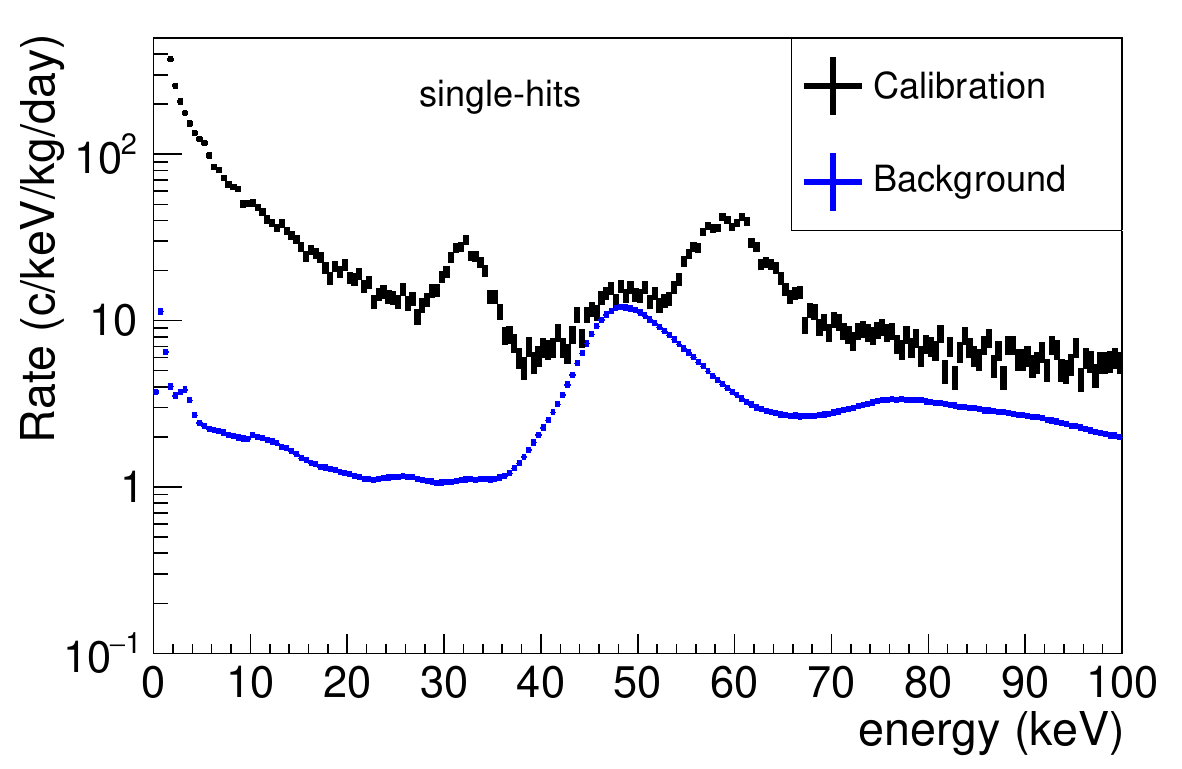}
  	  \caption*{\label{fig:neucal2_sing}}
  	\end{subfigure}%
  	~ 
  	\begin{subfigure}[b]{0.5\textwidth}
      \centering
  	  \includegraphics[width=\textwidth]{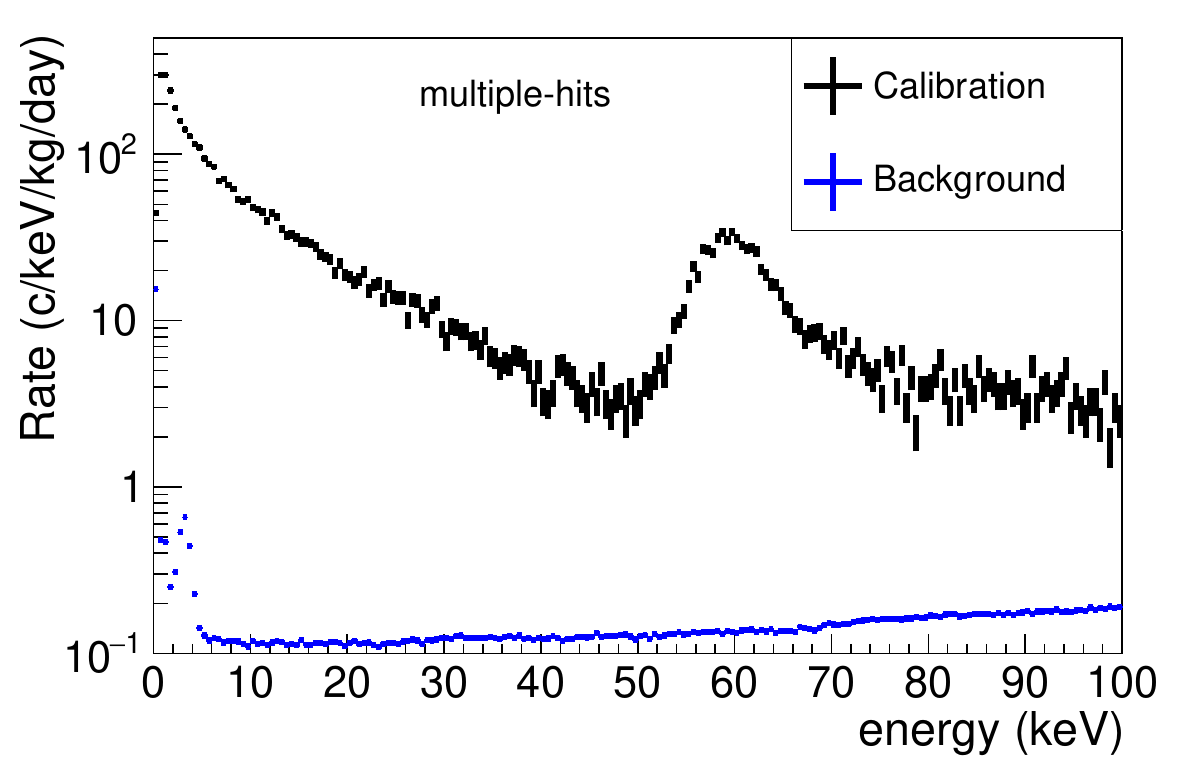}
  	  \caption*{\label{fig:neucal2_mult}}
  	\end{subfigure}%
    \vspace{-0.6cm}
    
    \caption{Energy spectrum produced at low energy in one of the \Cf calibrations (west side) of the \ANAIS experiment. Horizontal scale is represented in electron-equivalent energies. The background contribution is shown in blue as reference. Left panel: single-hits; right panel: multiple-hits.}
    \label{fig:neucal2}
\end{figure}

GEANT4~\cite{GEANT4:2002zbu} simulations of the neutron calibration runs at the \ANAIS set-up, including a detailed description of all the shielding and detectors components, allow concluding that events in the ROI correspond to nuclear recoils and were able to reproduce qualitatively the measured energy spectra for single and multiple hits, shown in Figure~\ref{fig:neucal2}, in spite of the uncertainties in the knowledge of the QF required for the comparison of simulations and measurements. In fact, we are aiming at determining the QF for our crystals by a precise quantitative comparison between simulation and measurement. It can be observed in Figure~\ref{fig:neucal2} that background contribution is negligible --except in the \Pb region, at about 40-50 keV in the single-hits spectrum. Moreover, contributions from electron/gamma events are only relevant at precise energies: 57~keV (inelastic scattering in $^{127}$I, also observed in multiple-hits spectrum) and 31~keV ($^{128}$I decay after neutron capture in $^{127}$I). Below 20~keV in the single-hits and 40~keV in the multiple-hits spectra, rates are dominated by multiple elastic scattering on Na and I nuclei. 
%

\begin{figure}
  	\begin{subfigure}[b]{0.5\textwidth}
      \centering
  	  \includegraphics[width=\textwidth]{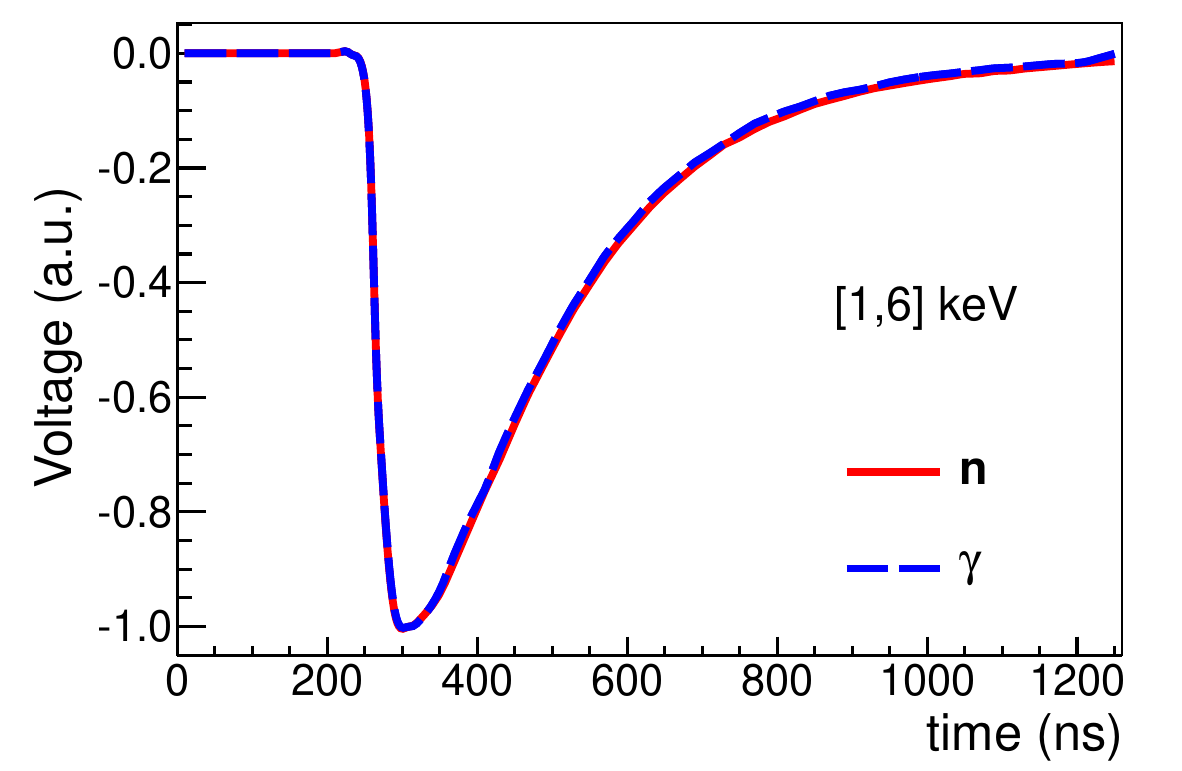}
  	  \caption*{\label{fig:neupulseROI}}
  	\end{subfigure}%
  	~ 
  	\begin{subfigure}[b]{0.5\textwidth}
      \centering
  	  \includegraphics[width=\textwidth]{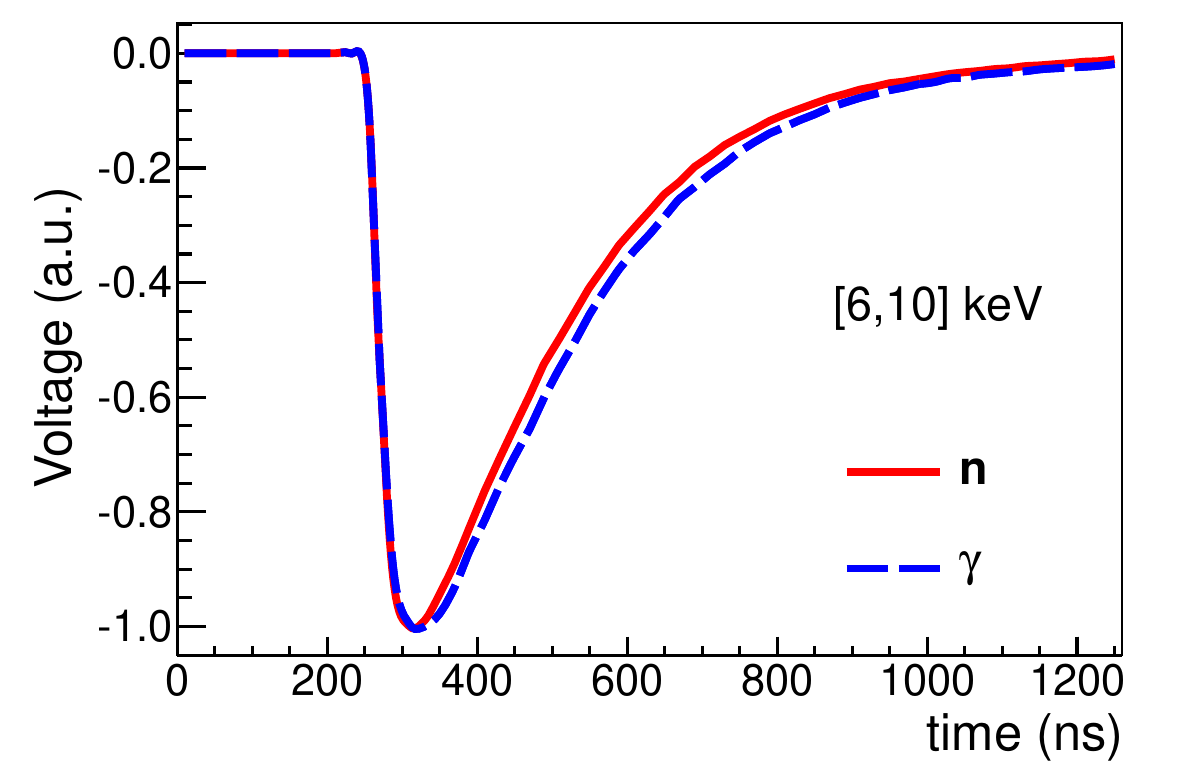}
  	  \caption*{\label{fig:neupulse610}}
  	\end{subfigure}%
    \vspace{-0.6cm}  	
     
    \caption{Average pulses corresponding to neutrons (solid red line) and gamma (dashed blue line) events for two energy regions: [1,6]~keV (left panel) and [6,10]~keV (right panel). Gamma events in the ROI have been selected in coincidence with a high energy $\gamma$, while gamma events outside the ROI correspond to single-hit background scintillation events.}
    \label{fig:neucal3}
\end{figure}

Figure~\ref{fig:neucal3} shows scintillation pulses observed for events corresponding to neutron interactions and electron/gamma events in the \ANAIS modules for two energy regions: [1,6]~keV (left panel) and [6,10]~keV (right panel). As it can be observed, in the ROI, nuclear recoil events (solid red line) show identical temporal behaviour than electron/gamma events (dashed blue line) selected in coincidence with a high energy $\gamma$ (\Na and $^{40}$K), whereas above 6~keV the former are slightly faster, as it was also observed in~\cite{Lee:2015iaa,KIMS:2018hch}. Both neutron interactions and \NaK coincidence events are occurring in the bulk of the crystal, homogeneously distributed, but the latter have a very low number of events. For this reason, we decided to train the BDT with nuclear recoil events in the \Cf calibration runs. Instead of using for the training events in the full ROI, from 1 to 6~keV, we decided to use only events from 1 to 2~keV, the range where our previous filtering showed to be weak. 
%


\subsection{The blank module set-up and performance}
\label{ssec:blank}

\begin{figure}
    \centering
    \includegraphics[width=0.5\textwidth]{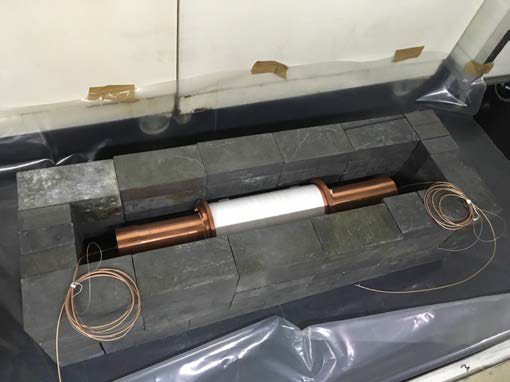}
    \caption{Setting-up the ANAIS blank module in the ANAIS hut, inside a specific lead shielding, 10~cm thick, next to ANAIS--112.}
    \label{fig:blank}
\end{figure}

A module similar to the \ANAIS modules, unless because of the absence of NaI(Tl) crystal, was integrated in the \ANAIS set-up in August 2018. We refer to this module as blank module. Two PMTs, identical to those used in the \ANAIS modules, were coupled to quartz optical windows at both sides of the module, consisting of a copper housing with the interior covered by Teflon diffusor film. This module was conceived to monitor specifically non-NaI(Tl) scintillation events that were identified as one of the main drawbacks limiting the sensitivity of ANAIS--112, starting the corresponding data taking with the beginning of the second year of operation of ANAIS--112. It is installed in the ANAIS hut at LSC, inside a specific lead shielding of 10~cm thickness, besides the \ANAIS set-up; the readout electronics, DAQ and analysis procedures applied are the same than for the \ANAIS modules. Figure~\ref{fig:blank} shows the set-up of this blank module. 
%

\begin{figure}
    \centering
   \includegraphics[width=0.9\textwidth]{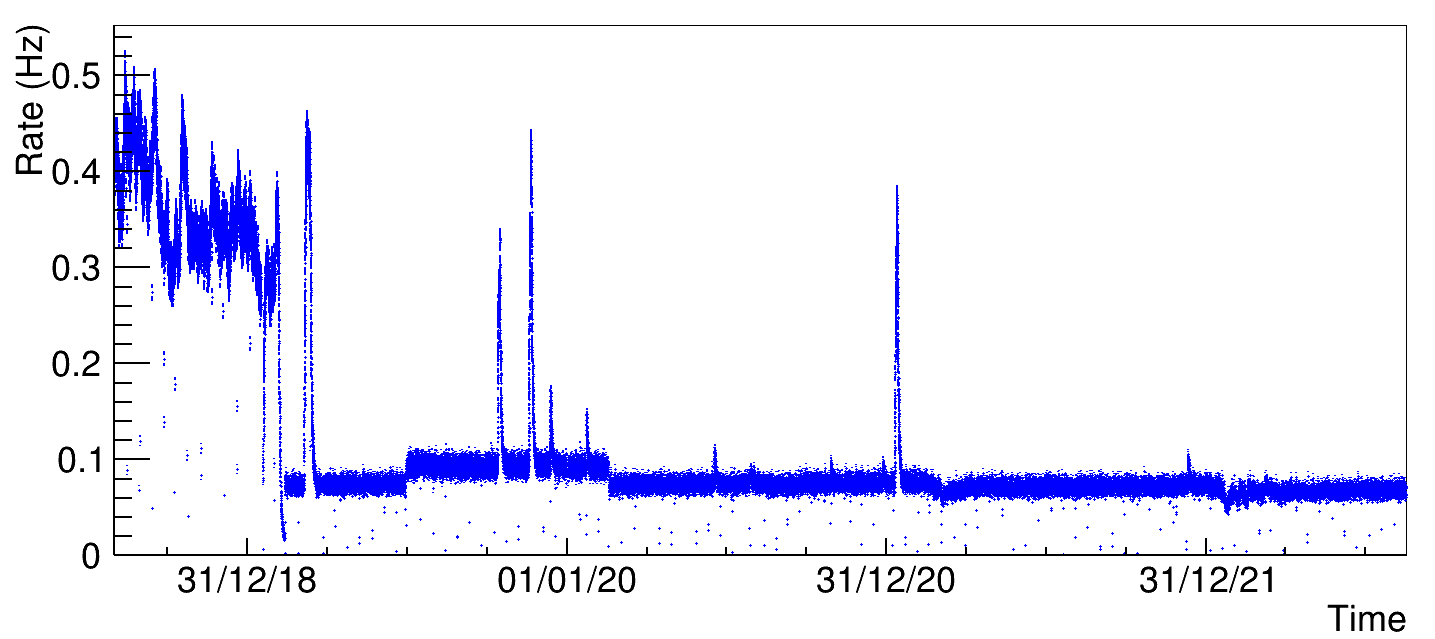}
    \caption{Trigger rate of the blank module corresponding to events with more than two peaks ($n_0+n_1$>2) after filtering for electric noise by removing events with high baseline RMS. Before February 2019, the air near the module had a mean radon content of 80~Bq/m$^3$. The strong decrease in rate observed corresponds to the flushing of the blank module shielding with radon-free air. The periods affected by shortcut of this radon-free air supply are clearly noticed as sharp peaks in the trigger rate. }
    \label{fig:blankrate}
\end{figure}

The module has been working uninterruptedly, accumulating a large amount of events of different origin. At the beginning of the blank module operation, the shielding was open to the laboratory air, which suffers from an average radon content of 80~Bq/m$^3$ but a high variability~\cite{Amare:2022dgr}. It was clearly observed that the detection rate of the blank module was sensitive to this radon content, observing a high correlation between both parameters in the period the blank module was exposed to normal air. In February 2019, the blank module shielding was tightly closed and flushed with radon-free air from the LSC supply and the detection rate of the blank module stabilised, as it can be seen in Figure~\ref{fig:blankrate}. The periods affected by shortcut of this radon-free air supply are clearly noticed as sharp peaks in the total trigger rate. The bias of the blank module PMTs was changed along the first year of operation in order to verify the right performance of the PMTs, and the trigger was tuned up. It can be concluded that radon in the air close to the PMTs is responsible of a relevant part of the coincident events, explaining trigger rates at the 0.5~Hz level. On a radon-free atmosphere, the trigger rate fell down to 0.1~Hz. It is worth to remark here that the \ANAIS modules are continuously operated in such a radon-free atmosphere. An upper limit of 0.06~Bq/m$^3$ at 95\% C.L. to the radon content was determined by screening the same nitrogen gas used for the flushing into the \ANAIS shielding using an HPGe detector at LSC~\cite{Amare:2022dgr}.
%

\newpage

\begin{figure}
    \begin{subfigure}[b]{0.5\textwidth}
      \centering
  	  \includegraphics[width=\textwidth]{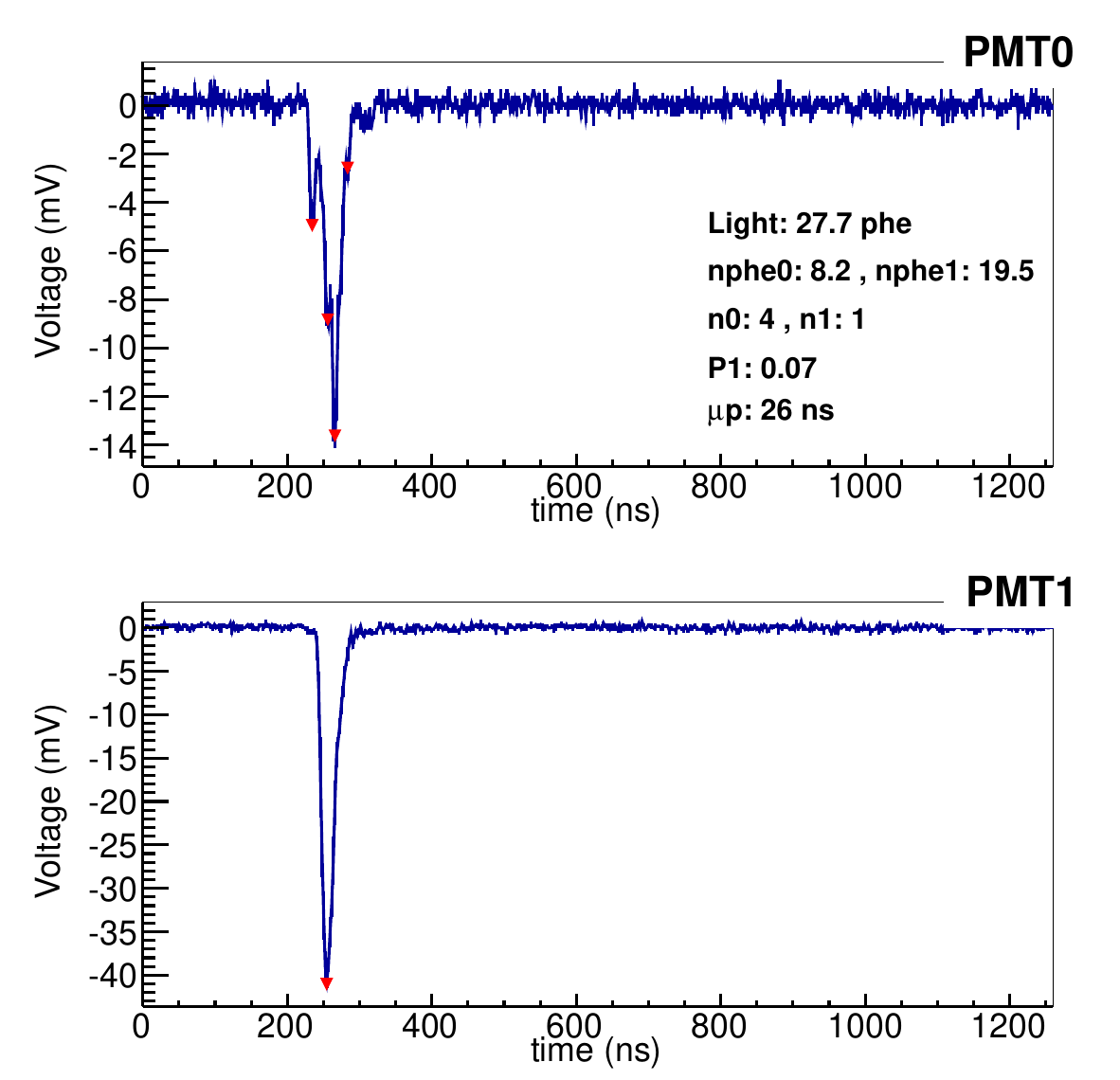}
  	  \caption{\label{fig:blankPulse1}}
  	\end{subfigure}%
  	~
    \begin{subfigure}[b]{0.5\textwidth}
      \centering
  	  \includegraphics[width=\textwidth]{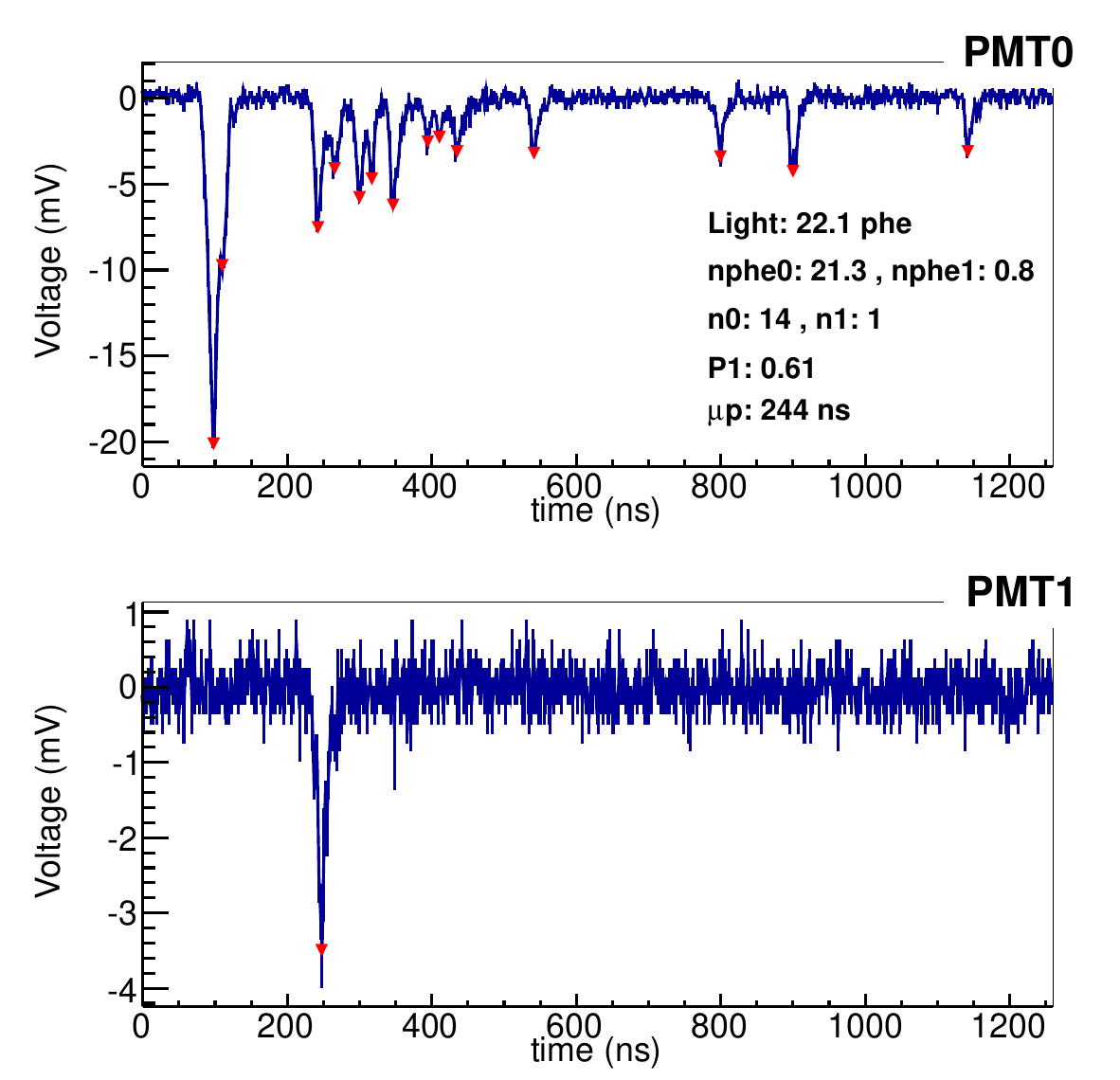}
  	  \caption{\label{fig:blankPulse2}}
  	\end{subfigure}%
  	
    \begin{subfigure}[b]{0.5\textwidth}
      \centering
  	  \includegraphics[width=\textwidth]{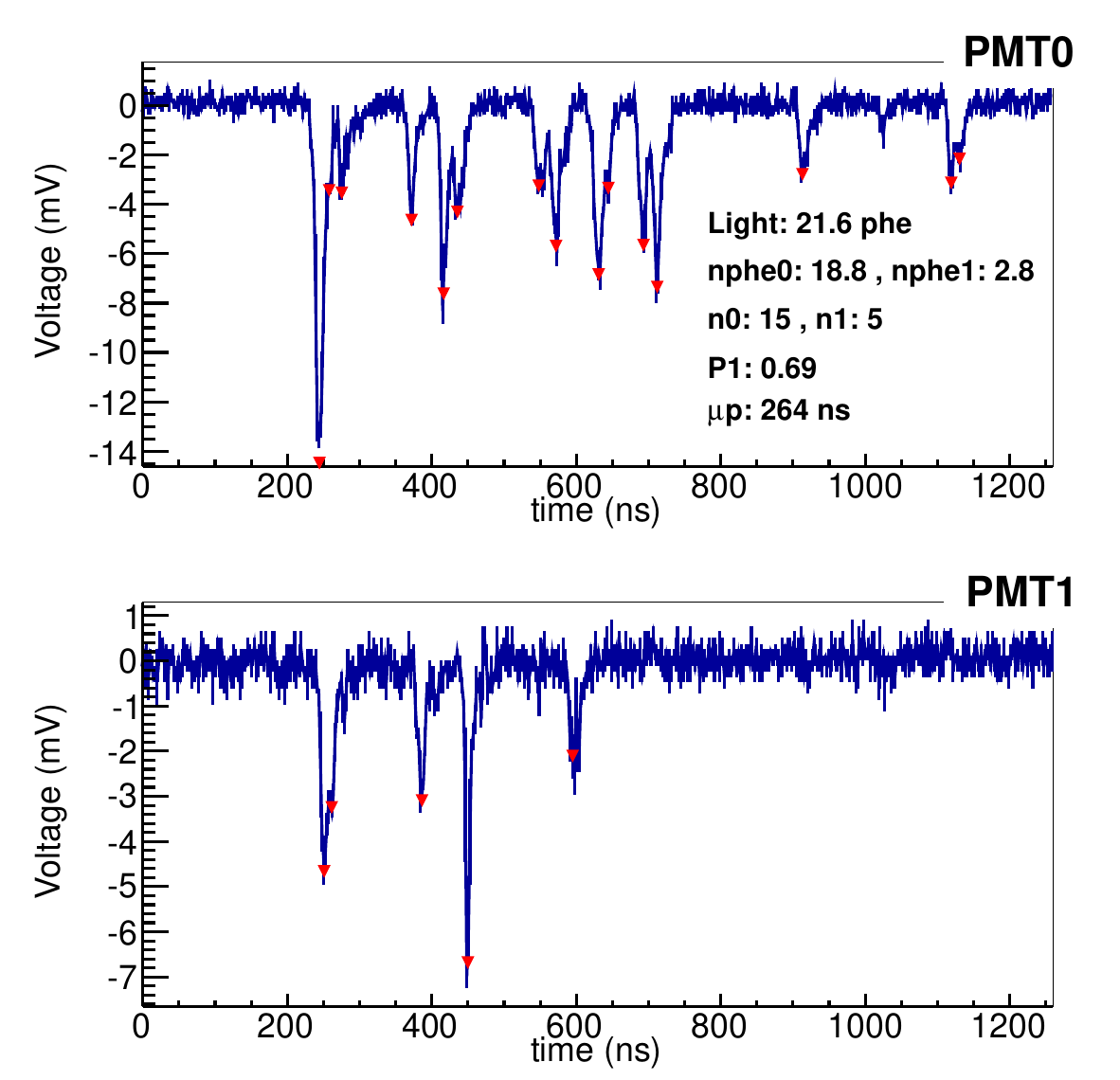}
  	  \caption{\label{fig:blankPulse3}}
  	\end{subfigure}%
  	~
    \begin{subfigure}[b]{0.5\textwidth}
      \centering
  	  \includegraphics[width=\textwidth]{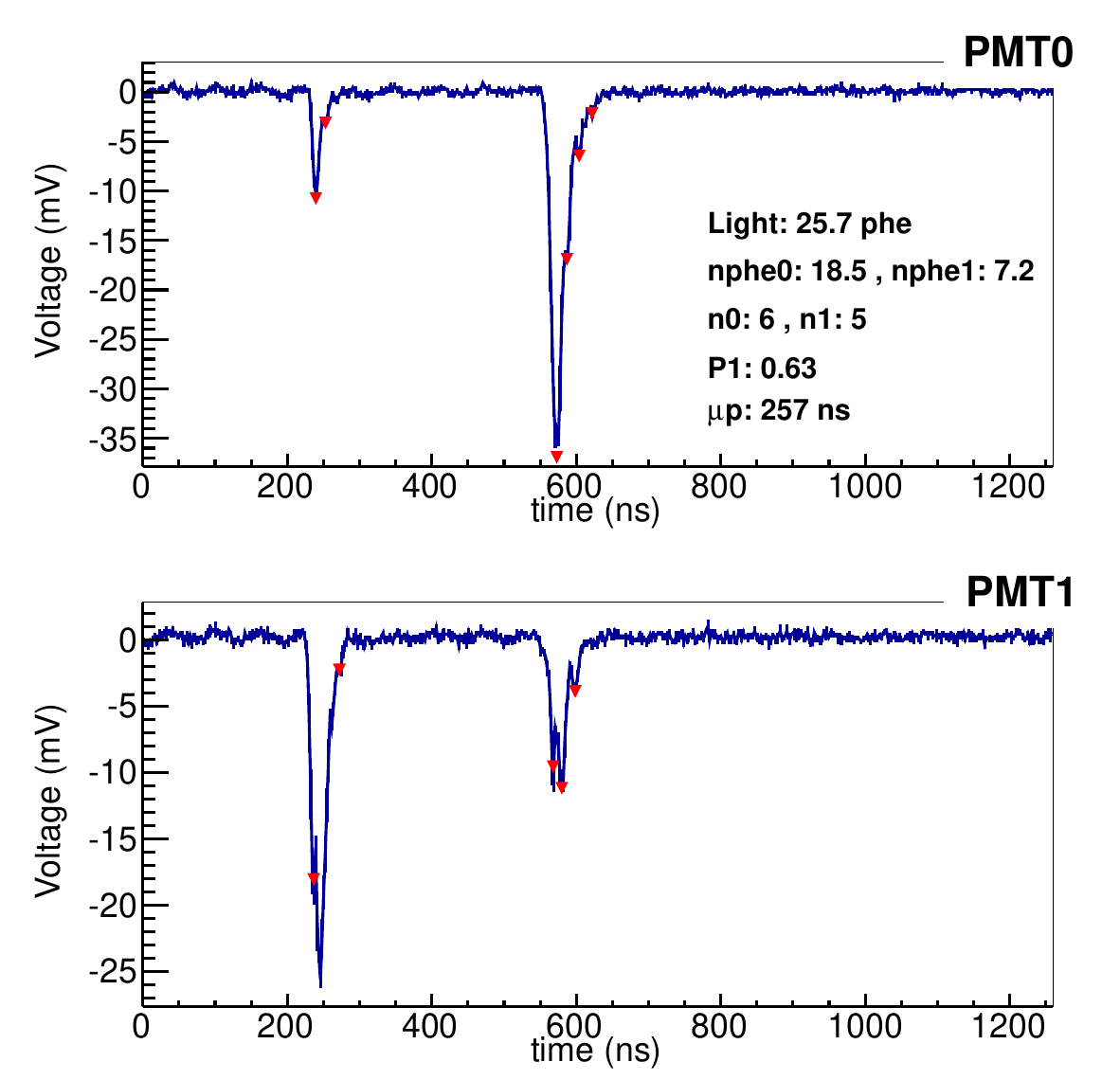}
  	  \caption{\label{fig:blankPulse4}}
  	\end{subfigure}%

    \caption{Examples of four events coming from the blank module. The two traces at each panel correspond to the two PMT signals, whereas the legend displays the total number of photoelectrons of the event and at every trace ($nphe_0$ and $nphe_1$), the number of peaks detected by the algorithm at every PMT waveform ($n_0$ and $n_1$), and pulse shape parameters $P_1$ and $\upmu_p$. Red triangles: peaks identified by the peak-searching algorithm. Most of the events detected are similar to that shown in the panel~(a), fast events with large amplitude, non-compatible with the NaI(Tl) scintillation shape. Other, like that shown in the panel~(b) are very asymmetric events, and are also easily removed by the filtering summarized in Section~\ref{ssec:filtering}. However, many other types of pulses have been identified that are not easy to discriminate from NaI(Tl) bulk scintillation, as those shown in panels~(c) and (d), which were not removed by that filtering. }
    \label{fig:blankpulses}
\end{figure}

Figure~\ref{fig:blankpulses} shows some examples of pulses coming from the blank module. Most of the events detected in this module are similar to that shown in the left-upper panel, fast events with large amplitude, non-compatible with the NaI(Tl) scintillation shape but with Cherenkov emission in the PMTs glass. Others, like that shown in the right-upper panel are very asymmetric events; both are easily removed by the filtering designed for \ANAIS data and explained in Section~\ref{ssec:filtering}. However, many other types of pulses have been identified that are not easy to discriminate from NaI(Tl) bulk scintillation, as those shown in the two lower panels, which were not removed by that filtering. This observation supported our goal of introducing these events in the design of a new filtering based on machine learning tools, explained in Section~\ref{sec:ML}. 
%

In order to introduce the blank module events in the \ANAIS analysis we had to assign them an equivalent energy, although a calibration similar to that of the \ANAIS modules was impossible. This is required for selecting those events that would appear in the ROI, for which the filtering should be optimized. For that goal, the number of photoelectrons in the pulse, $nphe$, is the convenient parameter. We convert the $nphe$ for blank module pulses into equivalent energy with the light collection corresponding to each \ANAIS module~\cite{anais2017LY,Amare:2021yyu}.


\section{Machine learning technique applied to \ANAIS data}
\label{sec:ML}

Machine learning algorithms have shown to be powerful tools in pattern recognition and decision taking. For our goal (the classification of events as signal or noise), decision trees have shown a high performance. They have been implemented for a similar goal in COSINE--100~\cite{COSINE-100:2020wrv} and SABRE~\cite{Calaprice:2022dxb} experiments, in both cases producing higher efficiencies and better rejection than traditional cuts. Other machine learning approaches are being applied in the direct detection of DM~\cite{Herrero-Garcia:2021goa}. Unsupervised machine learning methods, in particular, could offer powerful tools to identify anomalous events or unknown systematic errors.


\subsection{Boosted Decision Tree algorithm}
\label{ssec:alg}

A boosted decision tree (BDT) is a multivariate analysis technique that, taking into account the correlations between individual input parameters, allows several weak discriminating variables to be combined into a single powerful discriminator~\cite{Coadou2010,Coadou2022}. Decision trees are an extension of the commonly used cut-based selection strategy in which each event is analysed by means of pass/no pass decisions taken on a single variable at a time that are iterated until a stopping criterion is achieved. In fact, most events do not have all features of either signal or noise\footnote{In general, {\it background} is used instead of {\it noise}. However, in this article {\it noise} has been preferred to avoid confusion with radioactive background.}, therefore, decision trees do not reject right away events that fail a criterion, but check whether other criteria may help to classify these events properly. This helps to increase the acceptance efficiency of the filtering with respect to cut-based selection strategy. 

\begin{figure}
    \centering
    \includegraphics[width=0.45\textwidth]{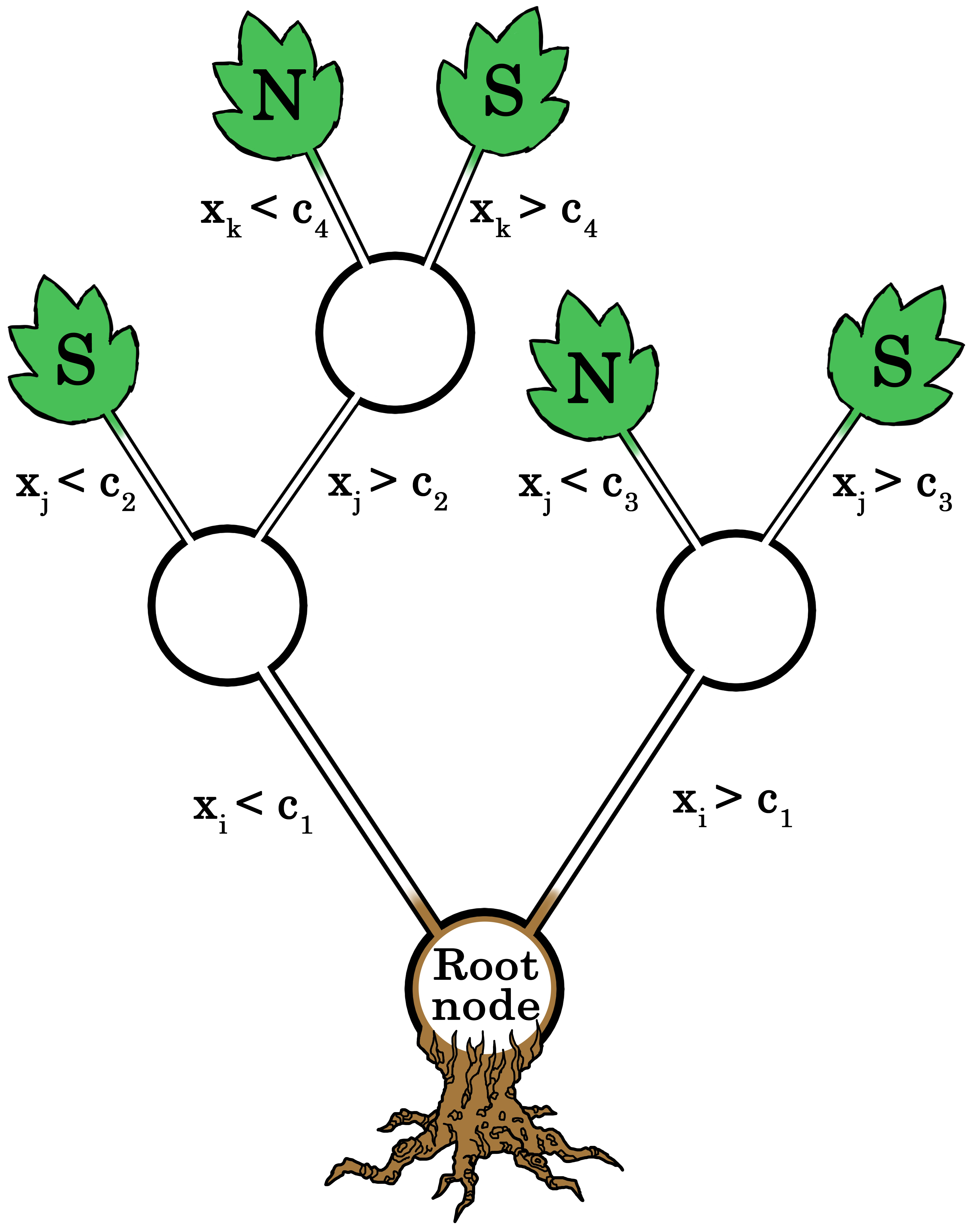}
    \caption{Schematic diagram of a decision tree with the architecture used in our analysis. Starting from the root node, a sequence of binary splits using the discriminating parameters $\vec{x}_i$ is applied to the data. Each split uses the parameter that at this node gives the best separation between signal and noise when being cut on (cut $c_p$). The same parameter may thus be used at several nodes, while others might not be used at all. The leaf nodes at the top end of the tree are labelled ``S'' for signal and ``N'' for noise depending on the majority of events that end up in the respective nodes.}
    \label{fig:BDTview}
\end{figure}

Mathematically, decision trees have a binary structure with two classes: signal and noise. In the training process, a sequence of binary splits using the discriminating parameters, $\vec{x}_i$, is applied to the training events in order to reach the best separation between signal and noise. Depending on the output value of the decision tree, the event is classified as signal or noise (see Figure~\ref{fig:BDTview}). The response of the decision tree with respect to fluctuations in the training sample can be stabilised using a boosting algorithm. The boosting of a decision tree~\cite{Freud1996} extends the aforementioned concept from one tree to several trees, forming a forest. The trees are derived from the same training ensemble by reweighting events, and are finally combined into a single classifier which is given by a weighted average of the individual decision trees. In particular, in this work we use AdaBoost (adaptive boosting)~\cite{Freud1996} to implement the boosting algorithm. Events that were misclassified during the training of a decision tree are given a higher event weight in the training of the following tree. In this way, subsequent iterations focus more on events misclassified by previous trees and a stronger classifier is built. Starting with the original event weights (=$1/N$, being $N$ the number of training events) when training the first decision tree, the subsequent tree is trained using a modified event sample where the weights of previously misclassified events are multiplied by a common boost weight derived from the fraction of misclassified events of the previous tree. The weights of the entire event sample are then renormalised such that the sum of weights remains constant. We adopt the purity, $p$, of the terminal node (or {\it leaf}) as a separation criterion. The purity is defined as $p=\frac{s}{s+n}$, where $s$ ($n$) is the sum of weights of signal (noise) events that ended up in this leaf during training. If $p>0.5$, +1 is attached to the leaf and is classified as signal, otherwise the leaf is set to $-$1 and labelled as noise. The trees are trained iteratively until a certain number of trees have been grown. If we denote the result of an individual classifier as $T(\vec{x}_i)$,  encoded for signal and noise as $T(\vec{x}_i)=+1$ and $-1$, respectively, adaptive boosting yields a response defined by:
\begin{equation}
  BDT(\vec{x}_i) = \frac{1}{n_{Trees}}\sum_{j=1}^{n_{Trees}} \ln(\alpha_j)\cdot T_j(\vec{x}_i),
\end{equation}
where $n_{Trees}$ is the number of trees in the forest, $\alpha_j=\frac{1-f_j}{f_j}\geq1$ the $j^{th}$ boost weight and $f_j$ the fraction of misclassified events by the $j^{th}$ decision tree. Small (large) values for $BDT(\vec{x}_i)$ indicate a noise-like (signal-like) event.
%

For the implementation of the BDT algorithm in ANAIS--112, we use the Toolkit for Multivariate Data Analysis (TMVA) package~\cite{TMVA2009} integrated in ROOT.


\subsection{Training populations}
\label{ssec:pop}
 
As commented before, training populations for signal and noise are obtained from the neutron calibrations using \Cf carried out in the last operation years of the \ANAIS experiment and blank module events, respectively. The most challenging aspect of a BDT training is to obtain pure event samples that are used to model the signal and noise events. For this reason, a series of preselection cuts on the training populations have been defined to extract representative samples of signal and noise. 

The neutron training population consists of events meeting the following requirements:
\begin{itemize}
    \item Energy in the 1-2~keV range.
    \item Slow charge ($P_1$) parameter larger than 0.35 to remove PMT-origin fast pulses that are found in the neutron calibrations and could be, at least partially, related with Cherenkov emissions in the PMTs.
    \item Trigger time lower than 400~ns from the main acquisition trigger. The software trigger position of the \ANAIS triggered events is around 250~ns within the 1260~ns digitization time window. Nevertheless, due to the high multiplicity of neutron interactions, the moderation of the neutron energy and the propagation time of the neutron between detectors, some multiple events show a delayed software trigger position, resulting in the tail of the pulse not being completely recorded. The estimates for the energy and the other pulse shape related parameters of such events are not correct and, therefore, they are discarded.
\end{itemize}
For the blank module training events, the criteria are:
\begin{itemize}
    \item Number of photoelectrons in the 10-28 range, corresponding to an equivalent energy of 1-2~keV.
    \item Events having $P_1$<0.35 are forced to be at 20\% of the total training population, although this kind of events are dominant in the detection rate of the blank module. However, they are easy to filter, whereas other events populations identified in the blank module require a training boost.
    \item Root mean square (RMS) of the baseline level below 0.60~mV in each PMT trace. Some of the blank module events correspond to electrical noise which can be removed based on the RMS of the baseline.
\end{itemize} 

Similar size training populations have been selected for signal and noise (above $3\cdot10^4$~events each). Moreover, only 70\% of the events of each population has been used for the training. The other 30\% has been used for testing in order to check the presence (if any) of overtraining (see Section~\ref{sec:valid}). 

Although ideally the training populations should be naturally clean, avoiding preselection cuts, to fully profit from the machine learning analysis potentiality, this is very difficult to achieve from the experimental point of view. The requirements imposed above for the training populations' selection correspond to ROI energy matching, signal quality cuts, and boosting of the noise population more difficult to reject, which do not imply any bias of the training population. Only the cut on the $P_1$ parameter on the neutron population is required to effectively clean it up from noise events because, even during a neutron calibration, events having PMT origin should be found besides the bulk scintillation events.

 
\subsection{Input parameters}
\label{ssec:vars}
 
Once the training populations for the scintillation and noise events have been determined, we define the input parameters for the BDT. Some of the discriminating parameters used in the BDT training are based on those best performing in the previous filtering protocol developed for \ANAIS and explained in Section~\ref{ssec:filtering} ($P_1$, $\mu_p$, $n_0$, $n_1$), but new parameters have also been added ($P_2$, $CAP$, $Asynphe$). These BDT training parameters are defined as:
\begin{enumerate}
  \item Slow charge ($P_1$): the fractional charge in [100, 600]~ns window with respect to total charge in [0, 600]~ns window for the total pulse,
  \begin{equation}
    P_1 = \frac{\sum_{100\textnormal{ ns}}^{600\textnormal{ ns}}A(t)}{\sum_{0\textnormal{ ns}}^{600\textnormal{ ns}}A(t)},
    \label{eq:P1}
  \end{equation}
  where $A(t)$ is the pulse amplitude at time $t$ after the software trigger position.
  
  \item Fast charge ($P_2$): the fractional charge in [0, 50]~ns window with respect to total charge in [0, 600]~ns window for the total pulse,
  \begin{equation}
    P_2 = \frac{\sum_{0\textnormal{ ns}}^{50\textnormal{ ns}}A(t)}{\sum_{0\textnormal{ ns}}^{600\textnormal{ ns}}A(t)}.
  \end{equation}
  
  \item Mean time ($\mu_p$): amplitude-weighted mean time of the arrival times of the peaks identified in the digitized window for the summed pulse, 
  \begin{equation}
   \mu_p = \frac{\sum_i A_it_i}{\sum_i A_i},
   \label{eq:mup}
  \end{equation}
  being $A_i$ and $t_i$ the amplitude and time of the $i^{th}$ peak identified in the pulse trace, respectively.
  
  \item Charge accumulated pulse ($CAP$): integral of the cumulative distribution of the total pulse charge up to a certain time, $x$, after the software trigger position,
  \begin{equation}
    CAP_x = \frac{\sum_{0\textnormal{ ns}}^{x\textnormal{ ns}}A(t)}{\sum_{0\textnormal{ ns}}^{t_{max}}A(t)},
  \end{equation}
  where $t_{max}$ is the end of the pulse. In particular, we use $x$ = 50, 100, 200, 300, 400, 500, 600, 700 and 800~ns.
  
  \item Asymmetry in the number of photoelectrons ($Asynphe$): the quotient between the partition in number of photoelectrons among both PMTs and the total number of photoelectrons. That is:
  \begin{equation}
    Asynphe = \frac{nphe_0-nphe_1}{nphe_0+nphe_1},
  \end{equation}
  where $nphe_0$ ($nphe_1$) is the number of photoelectrons in the pulse from PMT0 (PMT1).
  
  \item Number of peaks identified by the peak-searching algorithm in PMT0 and PMT1 traces ($n_0$ and $n_1$, respectively).
\end{enumerate}

\begin{figure}
  	\begin{subfigure}[b]{0.33\textwidth}
      \centering
  	  \includegraphics[width=\textwidth]{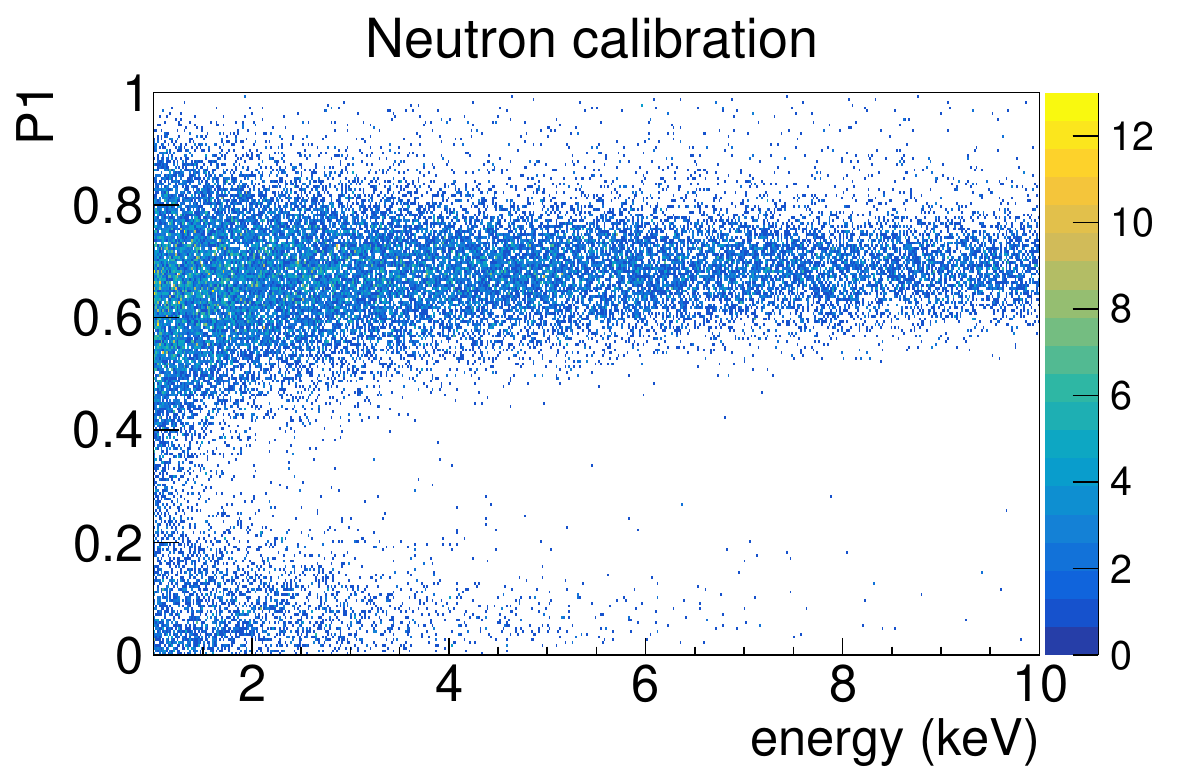}
  	  \caption*{\label{fig:P1LE_neu}}
  	\end{subfigure}%
  	~ 
  	\begin{subfigure}[b]{0.33\textwidth}
      \centering
  	  \includegraphics[width=\textwidth]{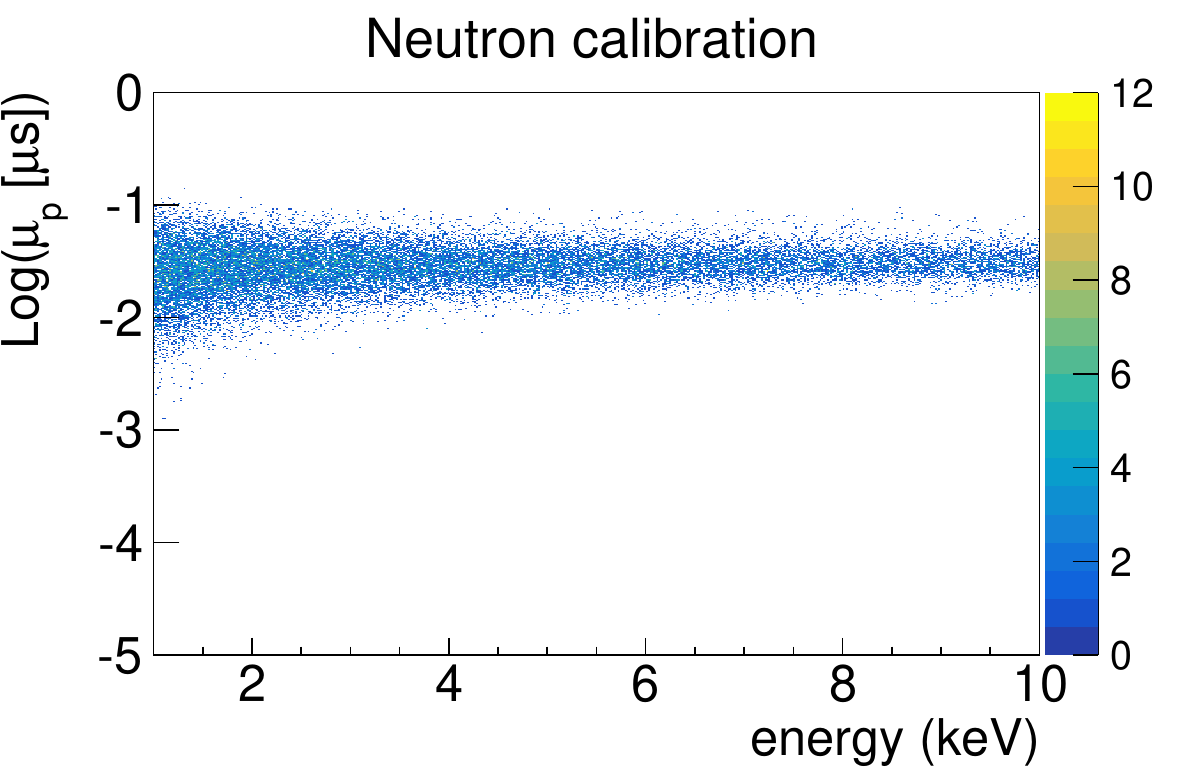}
  	  \caption*{\label{fig:FMLE_neu}}
  	\end{subfigure}%
  	~ 
  	\begin{subfigure}[b]{0.33\textwidth}
      \centering
  	  \includegraphics[width=\textwidth]{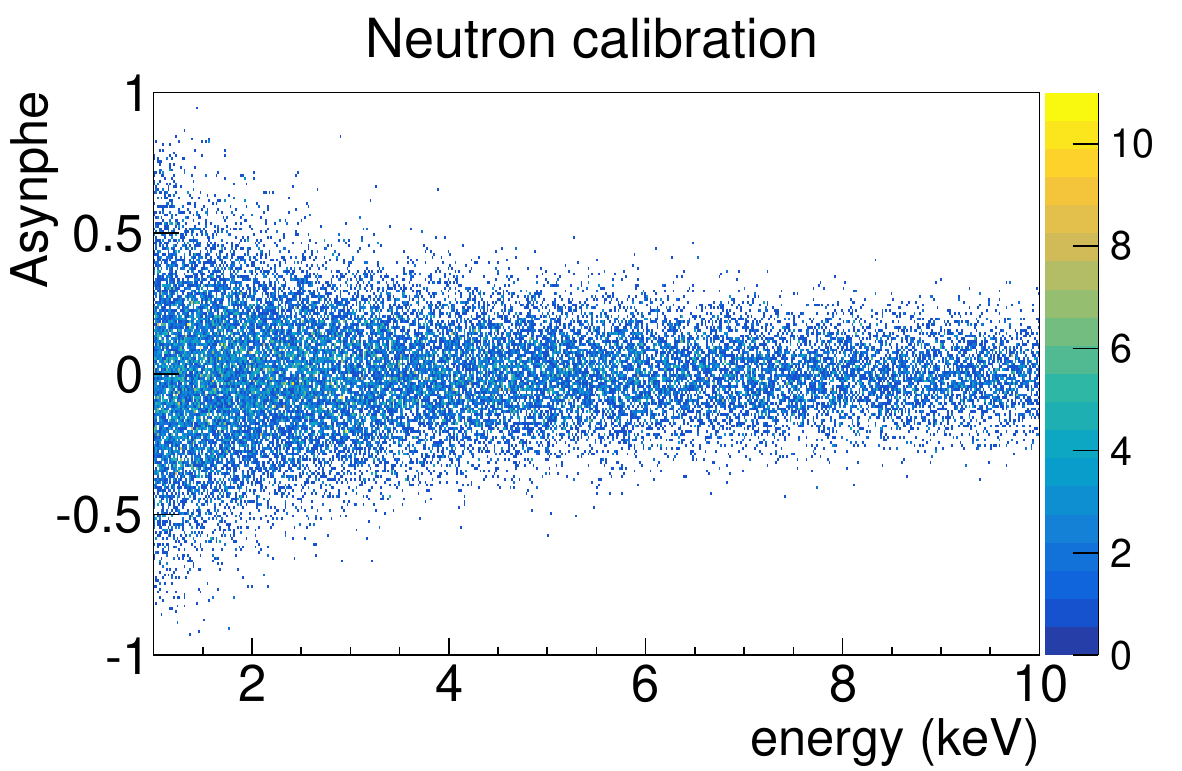}
  	  \caption*{\label{fig:AsyLE_neu}}
  	\end{subfigure}%
  	
  	\begin{subfigure}[b]{0.33\textwidth}
      \centering
  	  \includegraphics[width=\textwidth]{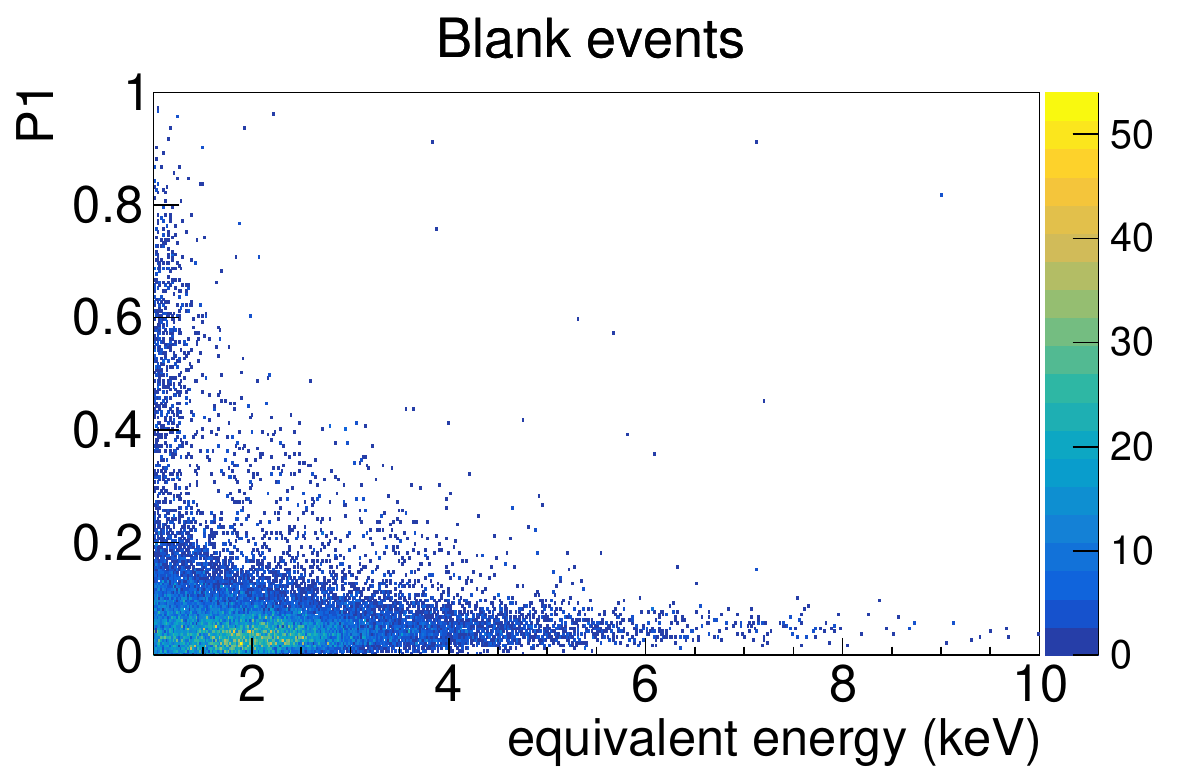}
  	  \caption*{\label{fig:P1Eeq_blank}}
  	\end{subfigure}%
  	~ 
  	\begin{subfigure}[b]{0.33\textwidth}
      \centering
  	  \includegraphics[width=\textwidth]{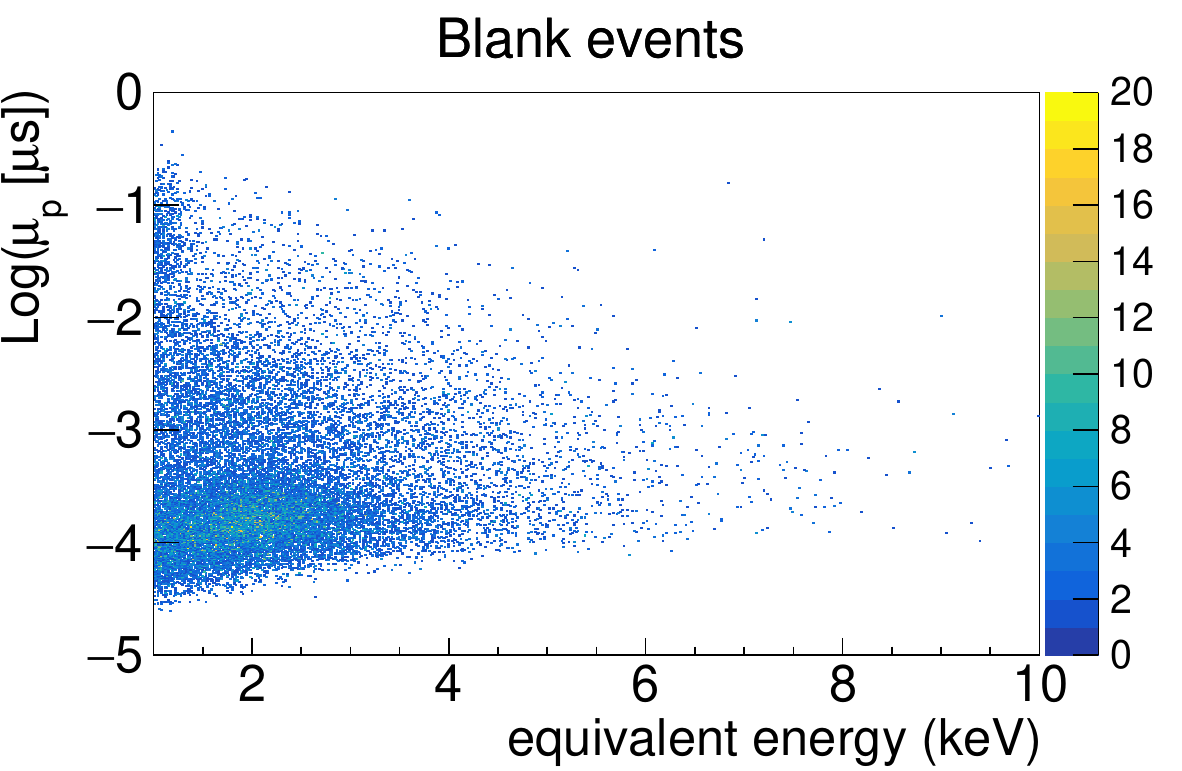}
  	  \caption*{\label{fig:FMEeq_blank}}
  	\end{subfigure}%
  	~ 
  	\begin{subfigure}[b]{0.33\textwidth}
      \centering
  	  \includegraphics[width=\textwidth]{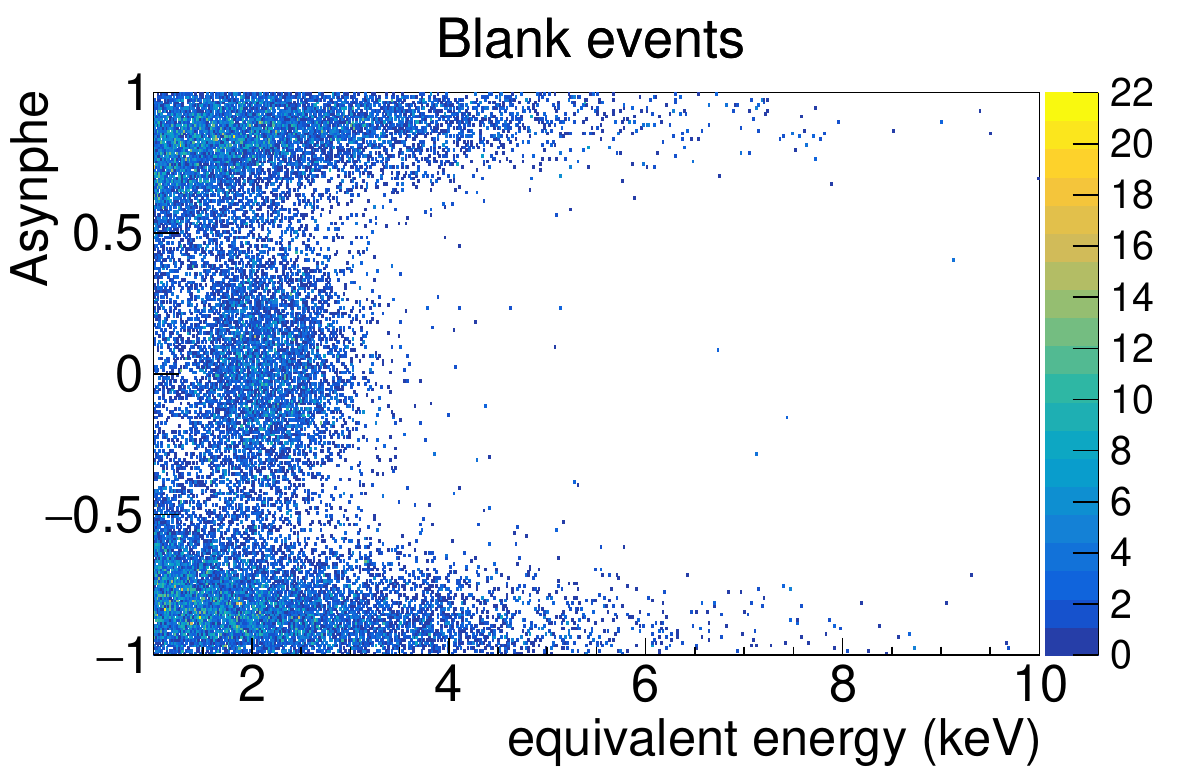}
  	  \caption*{\label{fig:AsyEeq_blank}}
  	\end{subfigure}%

  	\begin{subfigure}[b]{0.33\textwidth}
      \centering
  	  \includegraphics[width=\textwidth]{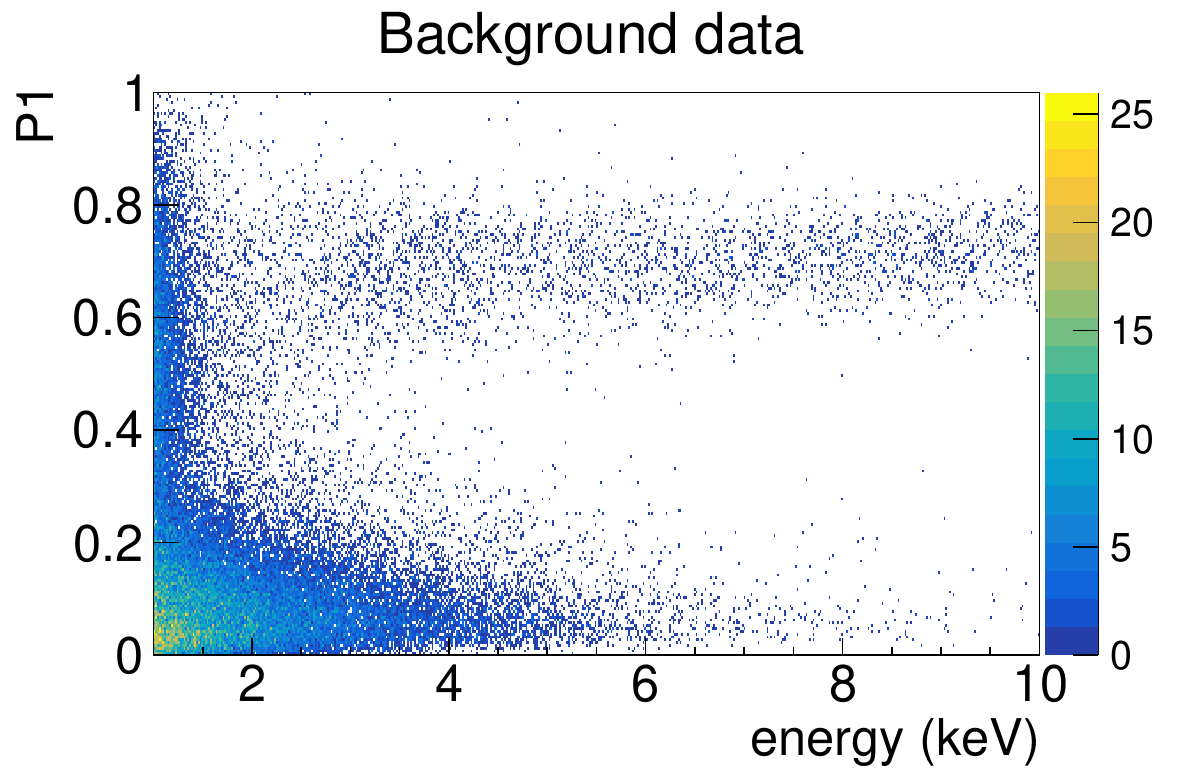}
  	  \caption{Slow charge, $P_1$ \label{fig:P1LE_bkg}}
  	\end{subfigure}%
  	~ 
  	\begin{subfigure}[b]{0.33\textwidth}
      \centering
  	  \includegraphics[width=\textwidth]{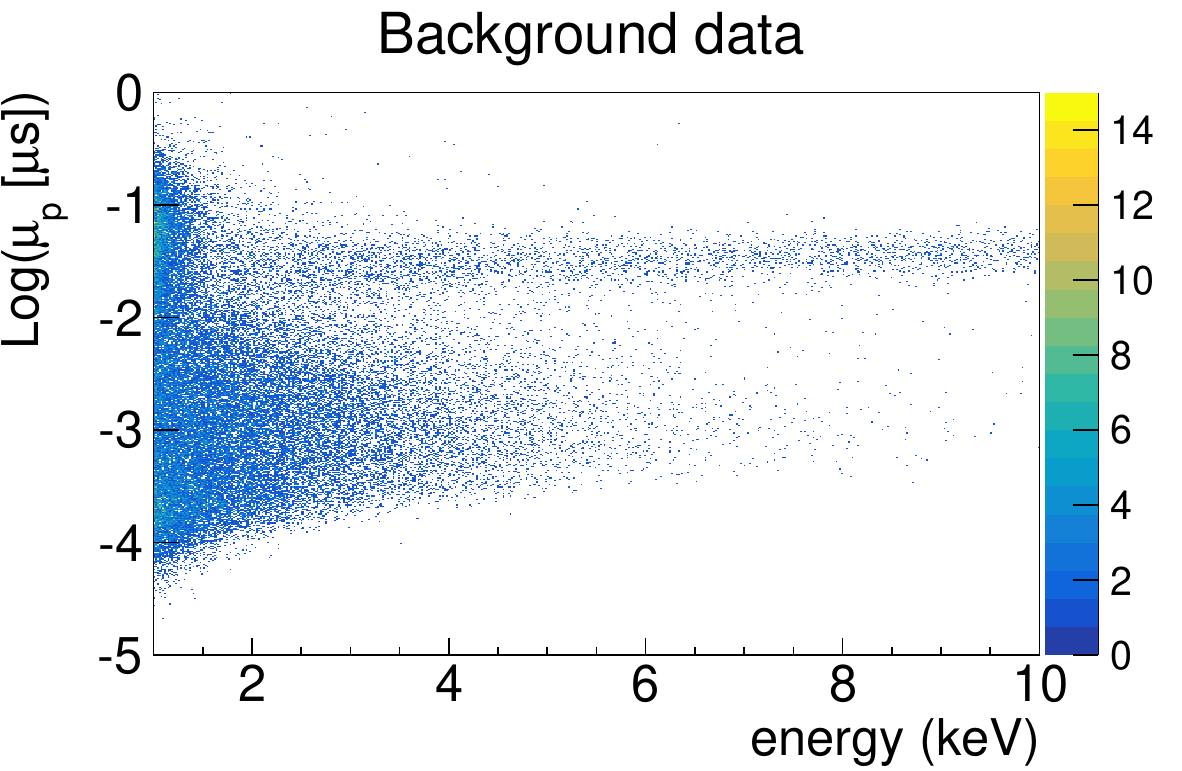}
  	  \caption{Mean time, $Log(\mu_p [\mu s])$ \label{fig:FMLE_bkg}}
  	\end{subfigure}%
  	~ 
  	\begin{subfigure}[b]{0.33\textwidth}
      \centering
  	  \includegraphics[width=\textwidth]{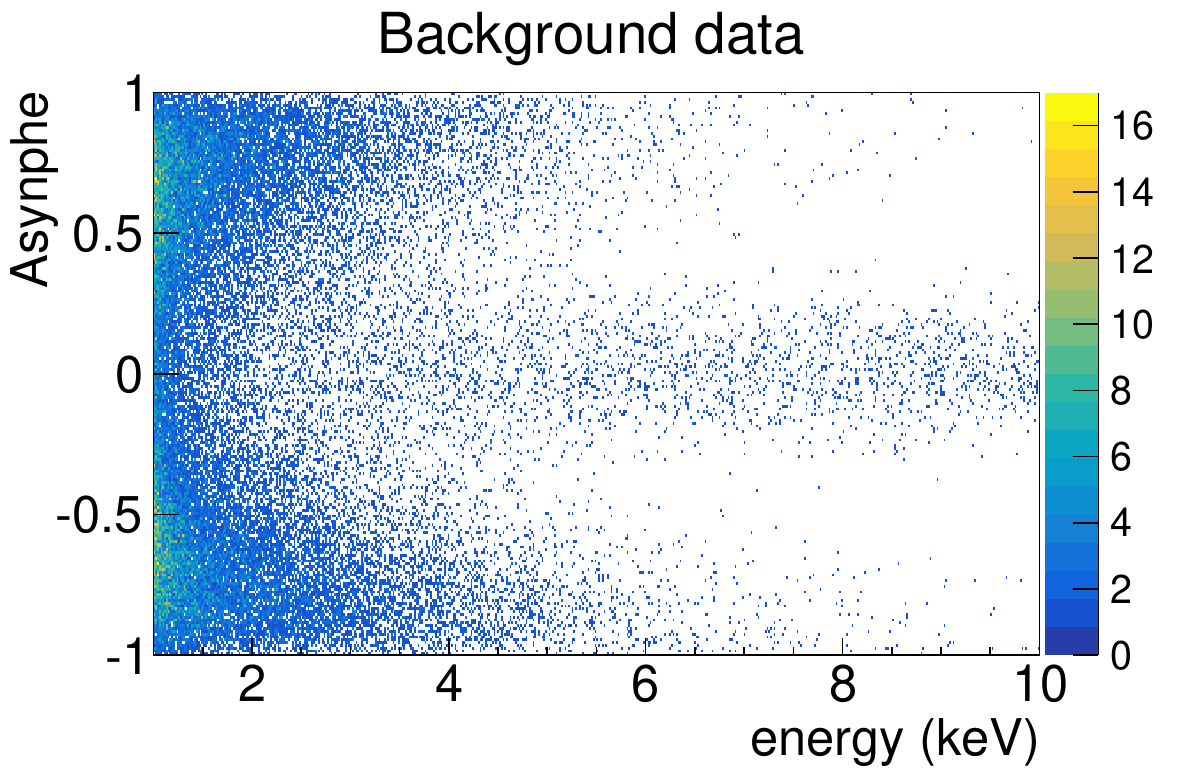}
  	  \caption{$Asynphe$ \label{fig:AsyLE_bkg}}
  	\end{subfigure}%

    \caption{Distribution of three representative input parameters as a function of energy from 1 to 10~keV: (a) Slow charge, $P_1$, (b) Mean time, $Log(\mu_p [\mu s])$ and (c) $Asynphe$. The upper panels show the distribution of events from neutron calibrations, whereas the lower panels display events from the background data. The same for events coming from the blank module with an equivalent energy is shown in the middle panels.}
    \label{fig:invarsLE}
\end{figure}

Figure~\ref{fig:invarsLE} shows the distribution of three representative BDT training parameters for the neutron population (upper panels) and the background data of the $\sim$10\% unblinded from the first year (lower panels) in the 1 to 10~keV energy region, and for events coming from the blank module with a number of photoelectrons equivalent to that energy (middle panels). The parameters shown are $P_1$~(a), $Log(\mu_p [\mu s])$~(b) and $Asynphe$~(c). For instance, considering the $P_1$ parameter, we can observe that the neutron population is slightly polluted by fast events with $P_1$<0.35 (especially at low energy) and, on the contrary, this kind of events are dominant in the detection rate of the blank module, as commented in Section~\ref{ssec:pop}.
%


\subsection{Training and BDT response}
\label{ssec:output}
 
Using the training populations and input parameters described above, we have carried out a 200-trees\footnote{The specific number of trees has been previously determined by scanning the value that minimizes the background level between 1 and 2~keV for the same training population. For other configurable parameters such as the maximum depth of the decision tree or the learning rate, the default values (3 and 0.5, respectively) have resulted in good performance.} training to develop a BDT score which can discriminate scintillation signal from PMT-noise more efficiently. All crystals are trained at the same time on a single BDT. Table~\ref{tab:rank} lists the separation signal/noise of the input parameters before training (2$^{nd}$ column) and the parameter ranking after training (3$^{rd}$ column). The ranking of the BDT input parameters is derived by counting how often the parameters are used to split decision tree nodes, weighting each split occurrence by the separation gain-squared it has achieved, and by the number of events in the node~\cite{TMVA2009}. The top parameter in the table is the best ranked. In particular, the parameters describing the asymmetry in the light sharing among both PMTs have the largest discriminating power ($n_0$, $n_1$, and $Asynphe$), followed by the new parameters defined on the integral of the total pulse. The other parameters used in the \ANAIS standard analysis ($P_1$ and $\mu_p$) have a discriminating power similar to the integrals of the pulse, but it should be noted that the discriminating capability of $P_1$ has been reduced by forcing the population of noise events coming from the blank module to be only lightly polluted by fast events (20\%, as explained in Section~\ref{ssec:pop}). 

\begin{table}
  \begin{center}
    \begin{tabular}{l c c}
    \toprule 
    Parameter & Parameter separation  & Parameter ranking \\
             & before training (\%) & after training (\%) \\
    \hline
    n$_0$       & 40.10 & 11.40 \\
    Asynphe     & 48.40 & 10.00 \\
    n$_1$       & 33.40 &  9.40 \\
    CAP$_{50}$  & 21.30 &  7.90 \\
    CAP$_{800}$ & 24.50 &  7.80 \\
    $\mu_p$     & 33.00 &  7.40 \\
    CAP$_{700}$ & 25.20 &  6.50 \\
    P$_1$       & 21.50 &  6.20 \\
    CAP$_{400}$ & 23.90 &  6.00 \\
    CAP$_{600}$ & 25.10 &  5.80 \\
    CAP$_{200}$ & 16.30 &  5.60 \\
    CAP$_{100}$ & 12.90 &  5.50 \\
    CAP$_{500}$ & 24.80 &  5.30 \\
    CAP$_{300}$ & 22.00 &  4.90 \\
    P$_2$       &  1.70 &  0.30 \\
    
    \toprule
    \end{tabular}
    
    \caption{Signal/noise separation of input parameters before training and parameter ranking after training. The top parameter is the best ranked.}
    \label{tab:rank}
  \end{center}
\end{table}

\begin{figure}
    \centering
    \includegraphics[width=0.65\textwidth]{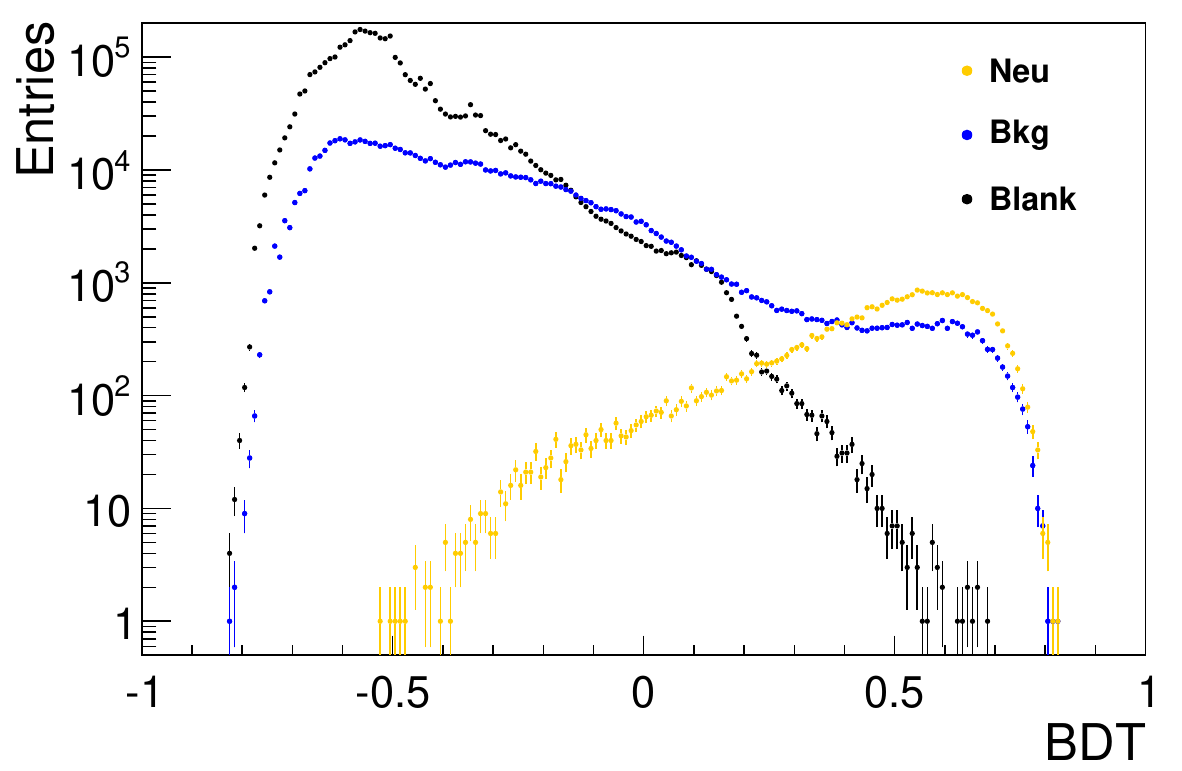}
    \caption{BDT distribution for neutron calibration events (in orange) and background events of the $\sim$10\% unblinded data (in blue) in the [1,2]~keV energy region. The same for events coming from the blank module with an equivalent energy is shown in black.}
    \label{fig:BDTdist}
\end{figure}

As a result of the training, we obtain the new BDT parameter that allows to better discriminate noise from scintillation events. Figure~\ref{fig:BDTdist} shows the BDT distribution for \Cf neutron calibration events (in orange) and background events of the $\sim$10\% unblinded data (in blue) in the [1,2]~keV energy region, and for events coming from the blank module with a number of photoelectrons equivalent to that energy (in black). We can observe that most of the events from the blank module are satisfactorily distributed around negative values of the BDT parameter and are negligible above 0.5. On the contrary, the events corresponding to neutron calibrations are distributed around positive values of BDT$\sim$0.60, although they present a tail of events that extends down to $-$0.5. The distribution of background events combines features of both populations, being dominated by noise-like events between 1 and 2~keV.
%


\section{Bulk scintillation events selection}
\label{sec:evsel}
\begin{figure}
  	\begin{subfigure}[b]{\textwidth}
      \centering
  	  \includegraphics[width=\textwidth]{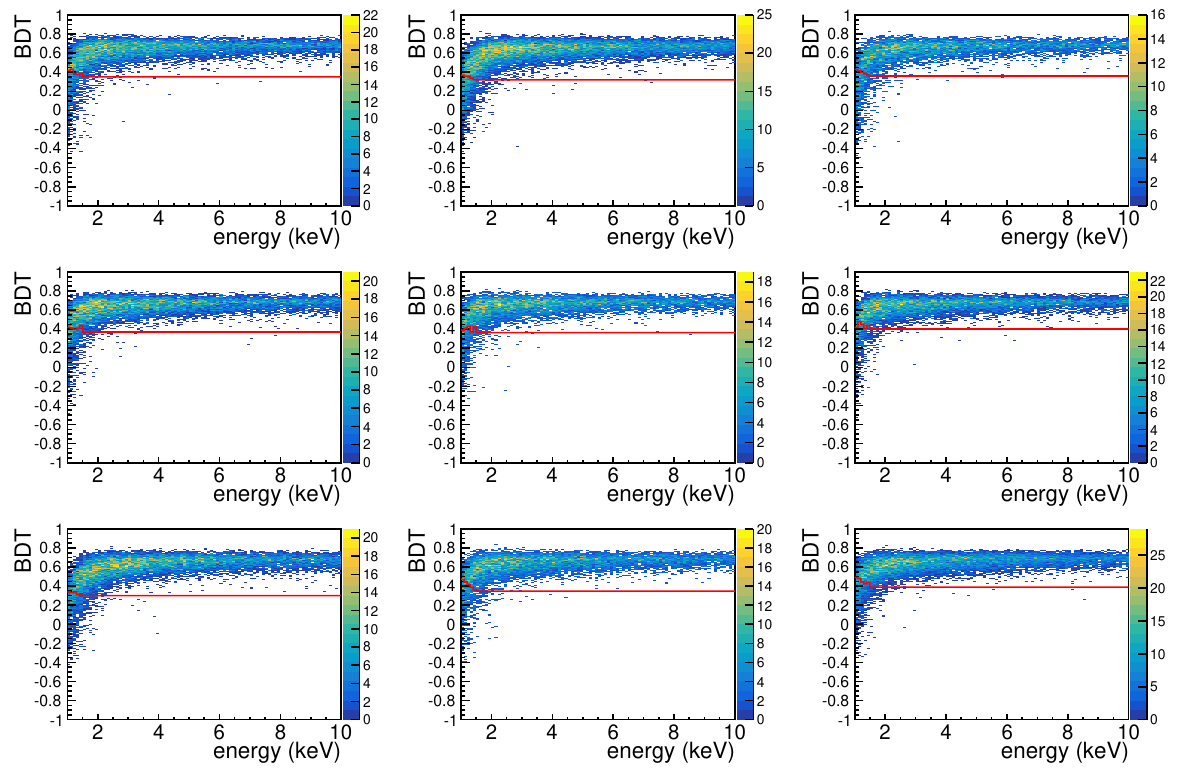}
  	  \caption{\label{fig:BDTLE_neu}}
  	\end{subfigure}%
  	
  	\vspace{0.3cm}
  	
  	\begin{subfigure}[b]{\textwidth}
      \centering
  	  \includegraphics[width=\textwidth]{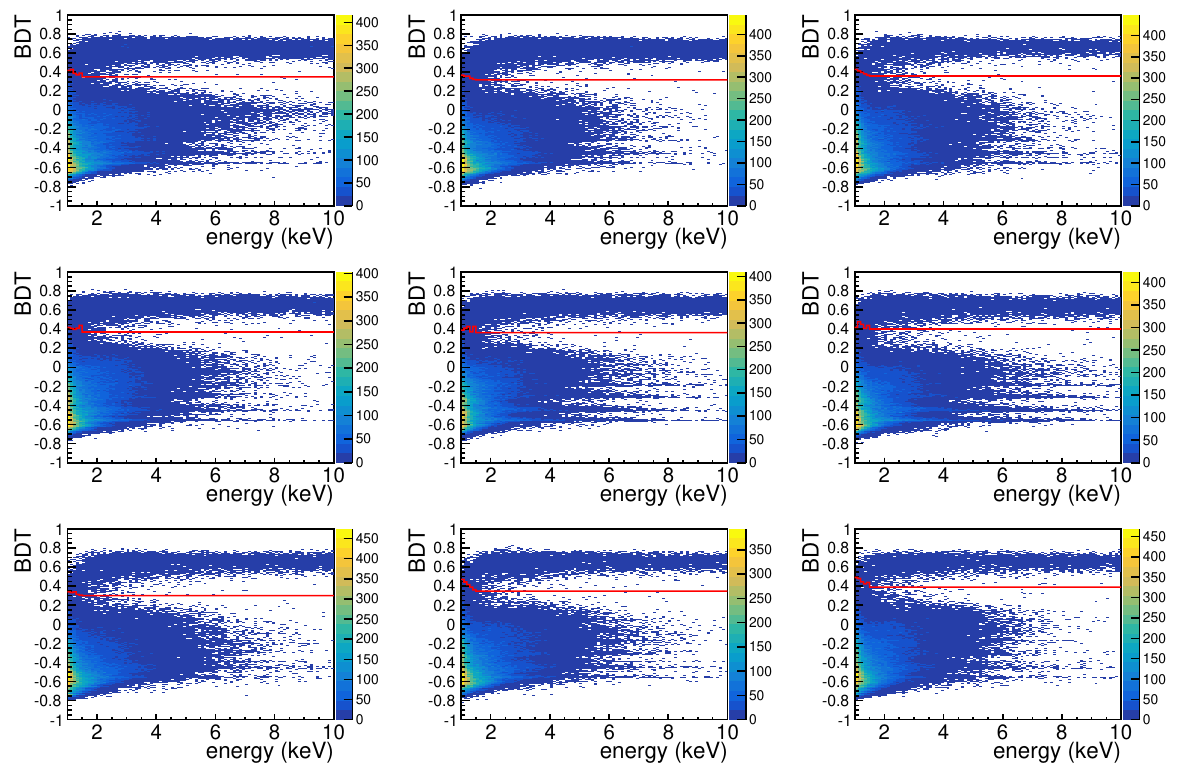}
  	  \caption{\label{fig:BDTLE_bkg}}
  	\end{subfigure}%

    \caption{BDT response as a function of energy for the nine modules for the neutron calibration sample (a) and for the $\sim$10\% unblinded dark matter search data (b). Solid red line: energy-dependent event selection cut.}
    \label{fig:BDTLE}
\end{figure}

\begin{figure}
    \centering
    \includegraphics[width=0.7\textwidth]{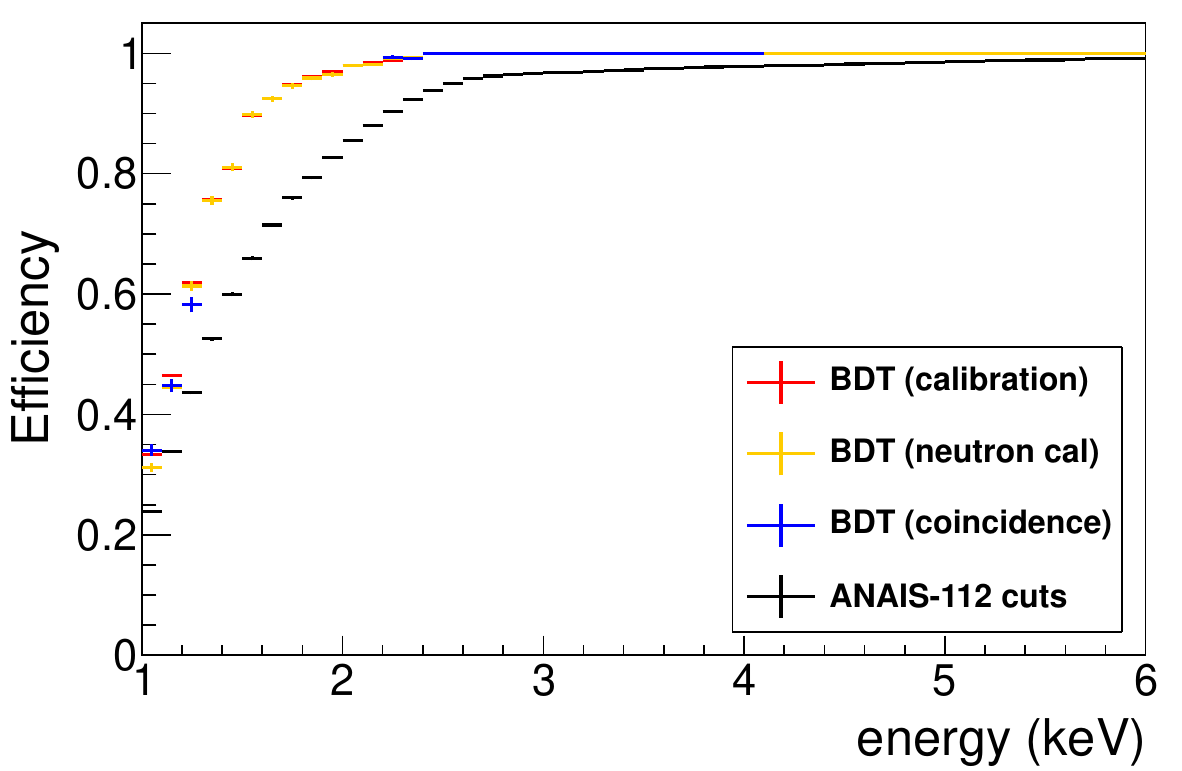}
    \caption{Average total efficiency obtained using the BDT cut estimated from neutron calibration (in orange), \Cd calibration (in red) and \NaK coincidences (in blue). Efficiency with the previous \ANAIS filtering protocols is shown in black for comparison.}
    \label{fig:eff}
\end{figure}

Figure~\ref{fig:BDTLE} displays the BDT response as a function of energy for the nine \ANAIS modules both for the neutron calibration sample~(a) and the $\sim$10\% unblinded background~(b). We can observe a clear separation between scintillation and fast PMT-noise events for energies greater than 2~keV, but below this energy the populations overlap. For the event selection, we define an energy-dependent BDT cut (solid red lines in Figure~\ref{fig:BDTLE}) and keep only events above it. This selection has been optimized for each detector and energy bin to ensure the lowest efficiency-corrected background from the $\sim$10\% unblinded data of the first year of data taking, guaranteeing acceptance efficiencies greater than those obtained with the current \ANAIS filtering procedures. The corresponding efficiency is estimated for each detector independently by using \Cf neutron calibration events. The ratio of the events which pass the signal selection to the total events is the acceptance efficiency\footnote{Following~\cite{anais2019Perf}, the efficiency estimated with the BDT cut has been multiplied by the trigger efficiency derived from the size of the coincidence window between both PMTs (200~ns).}. The average efficiency of the BDT cut is shown in orange in Figure~\ref{fig:eff}, while that obtained with the \ANAIS filtering protocols is displayed in black for comparison. We can observe that the acceptance efficiency derived from the BDT cut is significantly higher (around 30\% in [1,2]~keV, see Table~\ref{tab:bkg}) with respect to the previous \ANAIS filtering procedure. Furthermore, we use two additional populations to cross-check the consistency of the selection criteria and efficiency estimation: \Cd calibration single-hit events (in red), and the \NaK low energy events selected in coincidence with a high energy $\gamma$ (in blue). It has to be remarked that the latter efficiency has only been calculated around the 0.87 and 3.2~keV peaks from \Na and $^{40}$K, respectively, to avoid spurious coincident events, and that this efficiency estimate has a higher uncertainty due to the low number of \Na and \K coincidence events. It can be observed that all of them are well compatible.
%
%

\begin{figure}
    \centering
    \includegraphics[width=\textwidth]{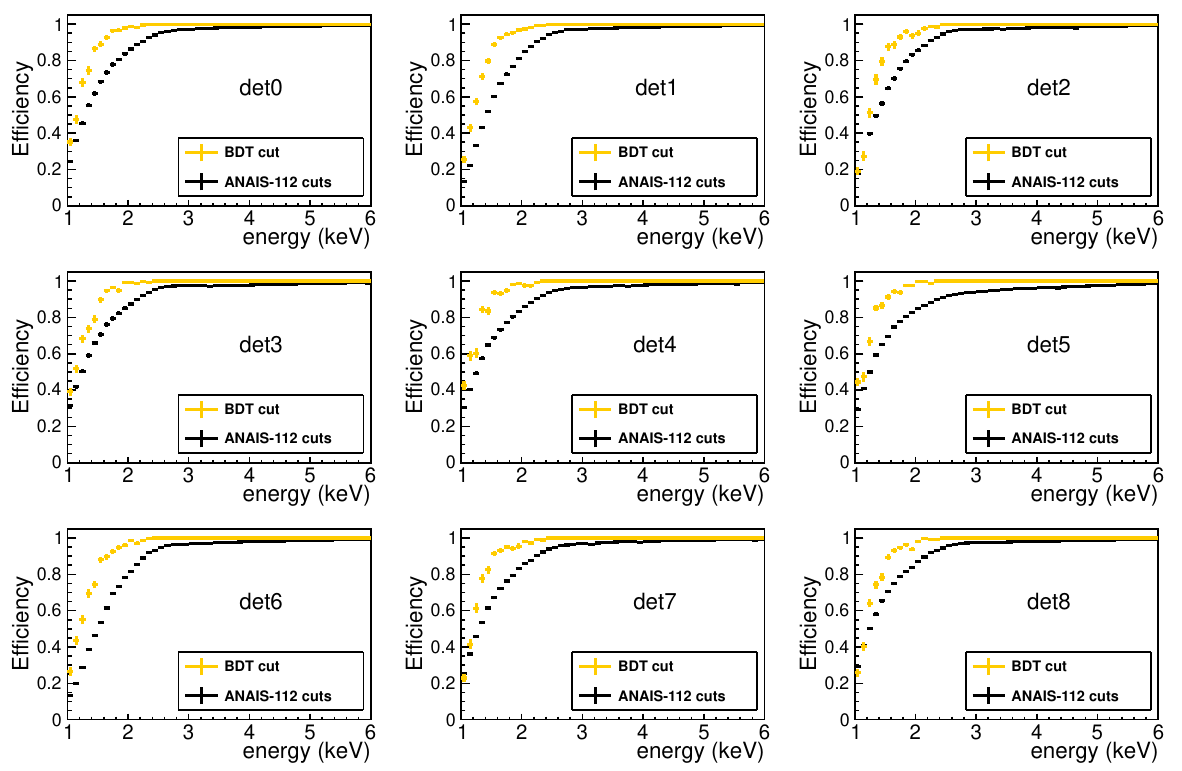}
    \caption{Total efficiency for each \ANAIS detector obtained using the BDT cut estimated from neutron calibrations (in orange). The efficiency with the previous \ANAIS protocols is shown in black for comparison.}
    \label{fig:eff3x3}
\end{figure}

In Figure~\ref{fig:eff3x3}, the efficiency derived from the BDT cut using \Cf neutron calibration events is shown in orange for each \ANAIS module. These efficiencies have clearly improved below 3~keV in all the modules with respect to those obtained with the previous \ANAIS filtering protocols (in black).
%

\begin{figure}
    \centering
    \includegraphics[width=\textwidth]{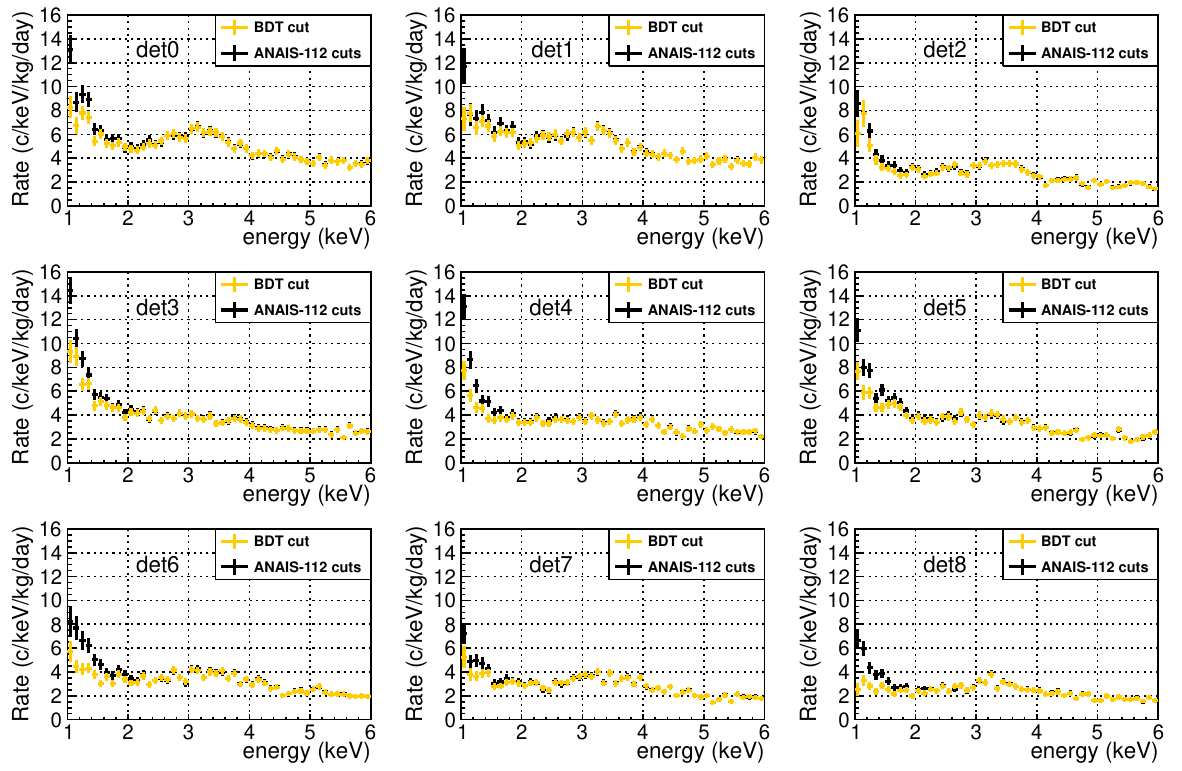}
    \caption{Anticoincidence energy spectrum measured in the ROI for each detector after BDT event selection and efficiency correction estimated from neutron calibration (in orange). In black, the same but using the previous \ANAIS filtering procedure. Data correspond to the $\sim$10\% data unblinded for the first year of operation.}
    \label{fig:bkg}
\end{figure}

The anticoincidence energy spectra in the ROI for the nine \ANAIS detectors after BDT event selection and efficiency correction are shown in orange in Figure~\ref{fig:bkg}. The background level at 2~keV ranges from 2 to 5~\ckkd, depending on the detector, and then increases up to 2--9~\ckkd{} at 1~keV. For comparison, the anticoincidence spectra using the previous \ANAIS filtering procedure are depicted in black in the figure. We can observe that the BDT method developed significantly reduces the background level below 2~keV for all detectors with respect to that obtained by the current \ANAIS procedure. In particular, the integral rate from 1 to 2~keV is 5.78$\pm$0.06 and 4.69$\pm$0.05~\ckkd{} for the \ANAIS filtering procedure and the BDT method, respectively, which represents a reduction of the background of 19\% (see Table~\ref{tab:bkg}).
%

\begin{table}
  \begin{center}
  \resizebox{15cm}{!}{
    \begin{tabular}{c ccc ccc}
    \toprule
    & & Measured rate & & & Detection efficiency & \\ 
    \cmidrule(r){2-4}
    \cmidrule(r){5-7}
    Detector & \ANAIS cuts & BDT cut & Variation  & \ANAIS cuts & BDT cut & Variation \\
             & (c/keV/kg/d) & (c/keV/kg/d) & (\%) & (\%) & (\%) & (\%) \\
    \hline
    D0        & 7.45$\pm$0.21 & 6.22$\pm$0.17 & -17 & 60.7$\pm$0.3 & 78.4$\pm$0.6 & 29 \\
    D1        & 7.30$\pm$0.24 & 6.48$\pm$0.18 & -11 & 52.2$\pm$0.3 & 74.4$\pm$0.5 & 43 \\
    D2        & 4.66$\pm$0.18 & 4.04$\pm$0.19 & -13 & 56.4$\pm$0.3 & 70.5$\pm$0.7 & 25 \\
    D3        & 7.12$\pm$0.20 & 5.92$\pm$0.17 & -17 & 64.2$\pm$0.3 & 78.6$\pm$0.6 & 22 \\
    D4        & 5.84$\pm$0.18 & 4.45$\pm$0.14 & -24 & 62.5$\pm$0.3 & 80.8$\pm$0.6 & 29 \\
    D5        & 6.20$\pm$0.18 & 5.07$\pm$0.15 & -18 & 63.0$\pm$0.2 & 80.4$\pm$0.6 & 28 \\
    D6        & 5.40$\pm$0.21 & 3.93$\pm$0.15 & -27 & 48.3$\pm$0.2 & 73.0$\pm$0.6 & 51 \\
    D7        & 4.21$\pm$0.15 & 3.55$\pm$0.14 & -16 & 60.1$\pm$0.3 & 75.5$\pm$0.6 & 26 \\
    D8        & 3.80$\pm$0.14 & 2.55$\pm$0.11 & -33 & 63.3$\pm$0.3 & 75.0$\pm$0.6 & 18 \\
    & & & \\ 
    \ANAIS    & 5.78$\pm$0.06 & 4.69$\pm$0.05 & -19 & 59.0$\pm$0.1 & 76.3$\pm$0.2 & 29 \\
    \toprule
    \end{tabular}
  }

    \caption{Measured rates after BDT filtering and efficiency correction in the [1,2]~keV energy region for each \ANAIS detector and on average, from the $\sim$10\% of unblinded data of the first year of data taking. The corresponding measured rates derived from the previous \ANAIS cuts are also presented for comparison in addition to the estimated percentage of variation. The last columns show the same for detection efficiency.}
    \label{tab:bkg}
  \end{center}
\end{table}


\section{BDT output validation}
\label{sec:valid}
 
\begin{figure}
  	\begin{subfigure}[b]{0.5\textwidth}
      \centering
  	  \includegraphics[width=\textwidth]{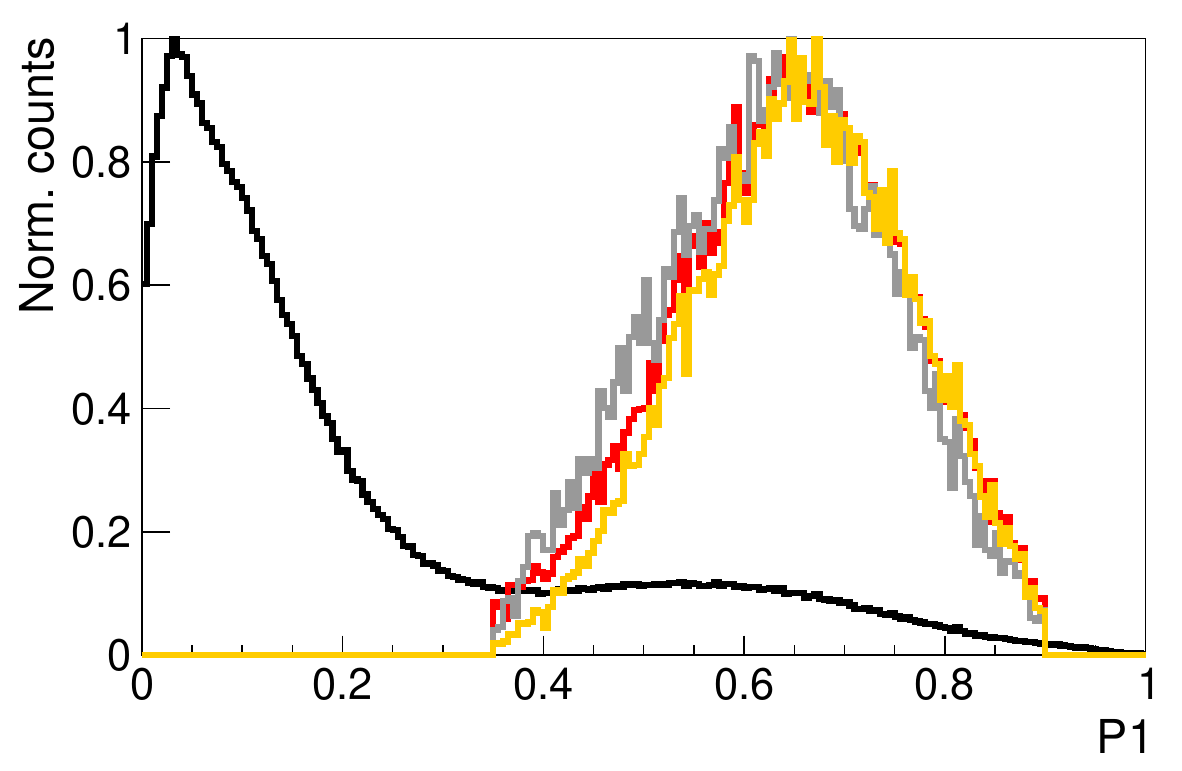}
  	  \caption*{\label{fig:App_P1}}
  	\end{subfigure}%
  	~ 
  	\begin{subfigure}[b]{0.5\textwidth}
      \centering
  	  \includegraphics[width=\textwidth]{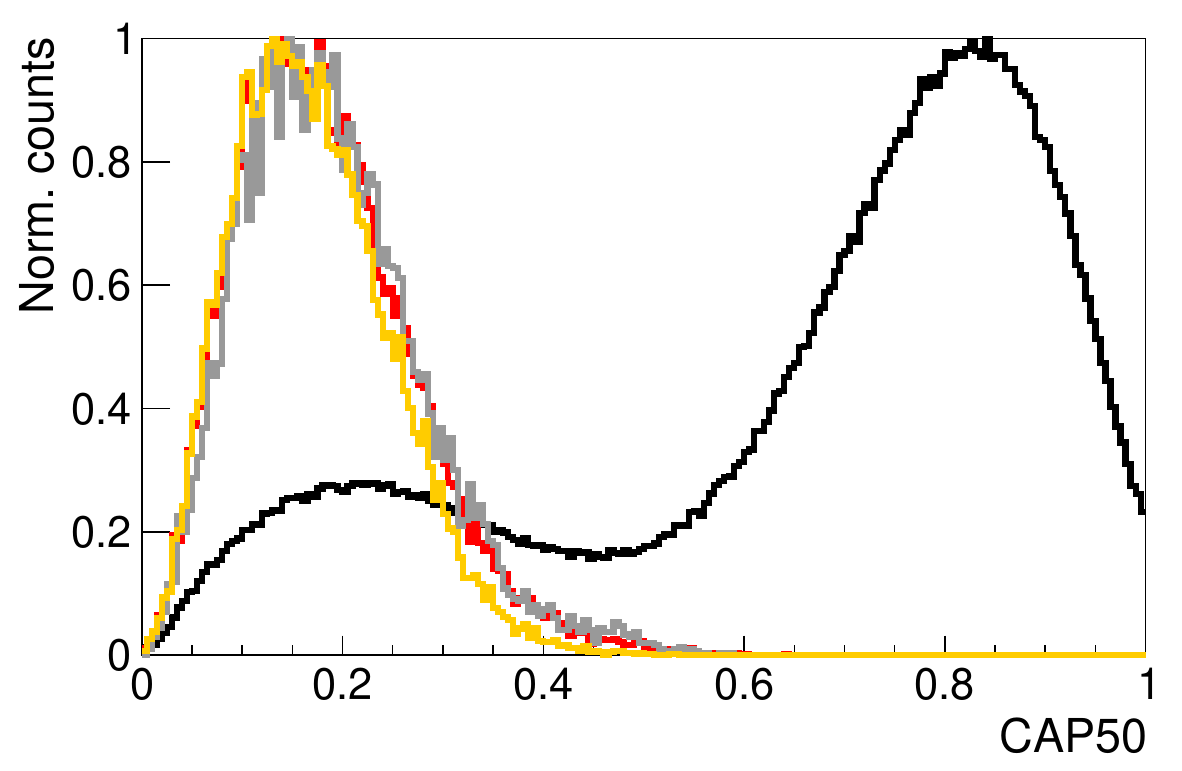}
  	  \caption*{\label{fig:App_CAP50}}
  	\end{subfigure}%

  	\begin{subfigure}[b]{0.5\textwidth}
      \centering
  	  \includegraphics[width=\textwidth]{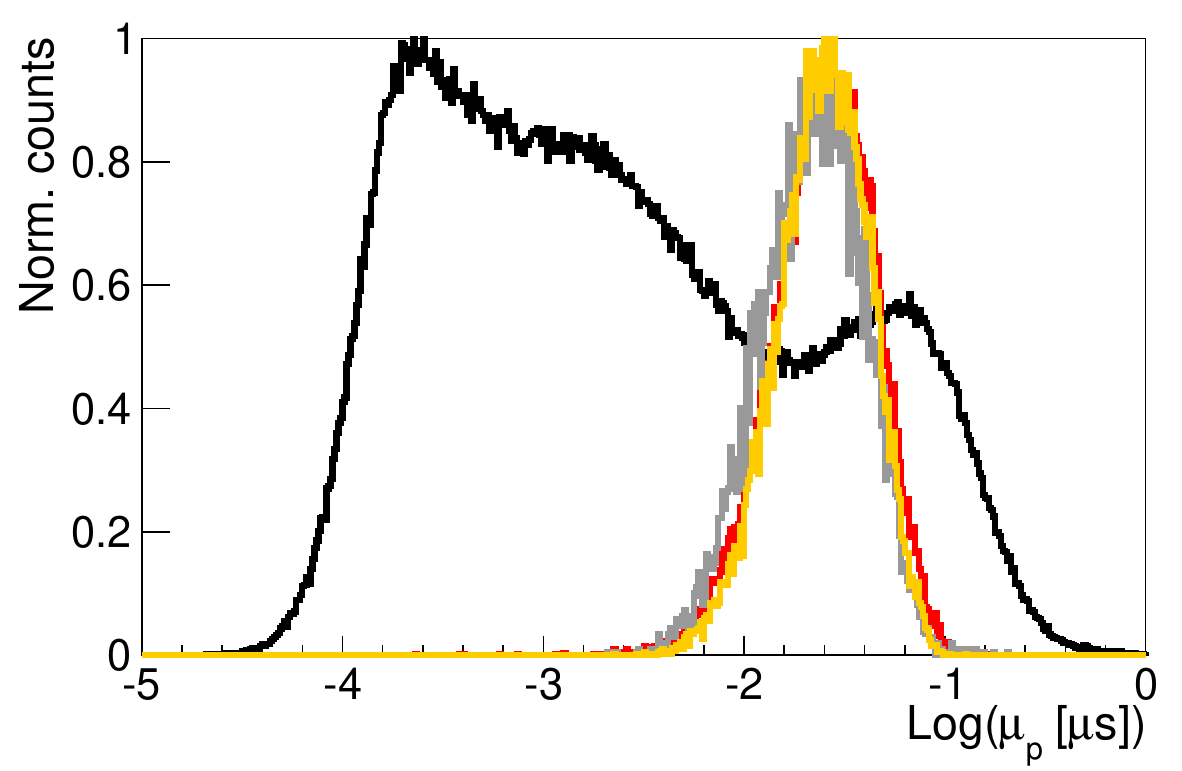}
  	  \caption*{\label{fig:App_FM}}
  	\end{subfigure}%
  	~ 
  	\begin{subfigure}[b]{0.5\textwidth}
      \centering
  	  \includegraphics[width=\textwidth]{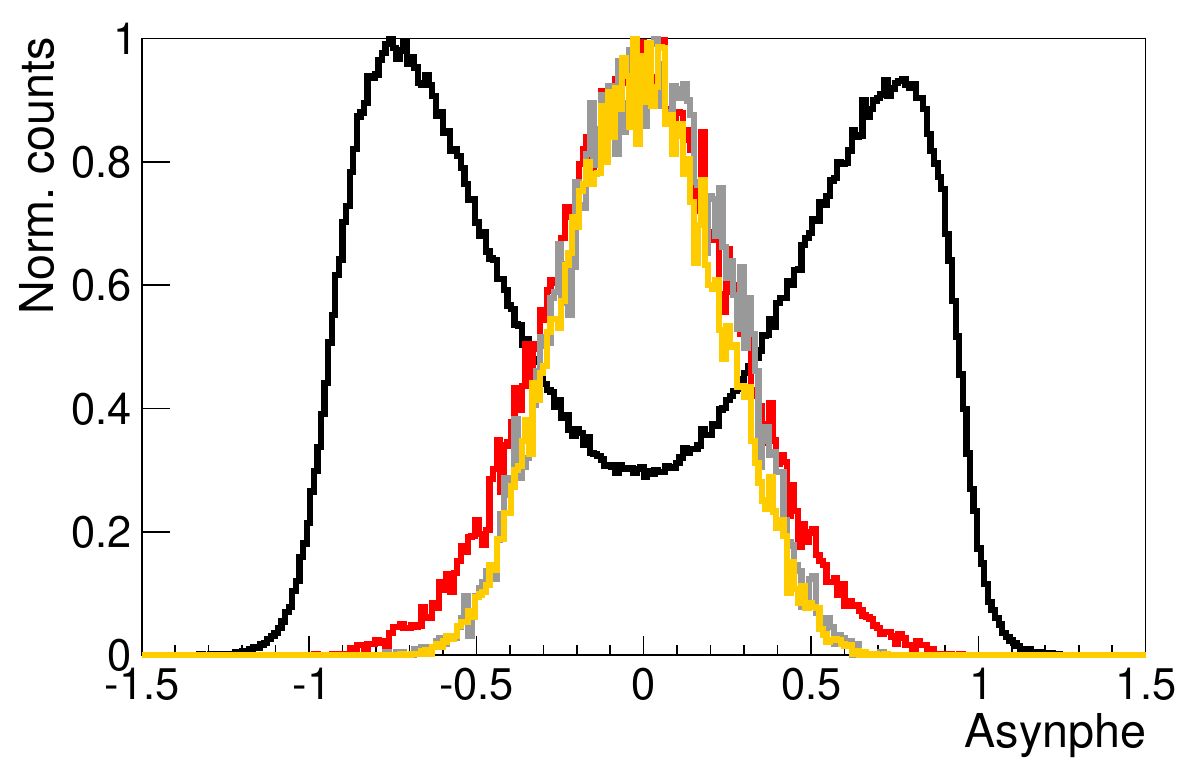}
  	  \caption*{\label{fig:App_Asynphe}}
  	\end{subfigure}%
  	\vspace{-0.6cm}

    \caption{Distribution of some representative input parameters used to validate the BDT output response. The black and red lines correspond to physics-run and \Cf neutron calibration data before the BDT selection, respectively. After applying the BDT criterion, scintillating-like events from the physics-run data are shown as gray line, whereas selected \Cf neutron calibration events are shown in orange.}
    \label{fig:App_invars}
\end{figure}

The BDT method provides a good event separation between scintillation and noise events, however, the validation of the algorithm becomes mandatory to ensure that neutron calibration events used in the training have the same behaviour as events corresponding to the background data. In order to do this, we compare the distribution of the input parameters for the neutron calibration training sample and for the events selected from the independent dark matter search data. Figure~\ref{fig:App_invars} displays the validation of some representative input parameters used to build the \ANAIS BDT. The black line corresponds to the raw background, whereas the red line is the preselected \Cf neutron calibration data. After the BDT event selection, scintillation-like events from the physics-run (\Cf neutron calibration) data are shown as gray (orange) lines. We can observe that there exists a good agreement in the shape of distributions for parameters between \Cf neutron calibration events and the background selected scintillation data. Similar results are obtained for the other parameters. The consistency between the two independent samples provides an indirect validation of the procedure.
%

We have also checked the possibility of overtraining in our BDT by comparing the efficiency curve obtained (for each detector), as shown in Figure~\ref{fig:eff3x3}, with those obtained by applying the training to the test population (see Section~\ref{ssec:pop}). Since the efficiency curves are compatible with each other, we can conclude that the BDT is not overtrained.


\section{\ANAIS sensitivity improvement with BDT filtering}
\label{sec:sens}

The \ANAIS sensitivity is determined by the standard deviation of the modulation amplitude estimator, $\sigma(\hat{S}_m)$~\cite{Coarasa:2018qzs}. Then, we quote our sensitivity to the annual modulation signal measured by the \DL collaboration at a given exposure as the ratio between the modulation amplitude measured by \DL in the corresponding energy region (0.0105$\pm$0.0011 and 0.0102$\pm$0.0008~\ckkd{} in [1,6] and [2,6]~keV, respectively~\cite{Bernabei:2018jrt}) and $\sigma(\hat{S}_m)$, $\Sen(\hat{S}_m,t) = S_m^{\text{DAMA}}/\sigma(\hat{S}_m)$, because it gives directly (in $\sigma$ units) the compatibility of the \DL signal with the null hypothesis ($S_m=0$) in ANAIS--112. To estimate the \ANAIS sensitivity, we exploit the Monte Carlo technique described in~\cite{ivanThesis}. The expected number of events for every time bin $t_i$ and detector $d$ can be written (following~\cite{Amare:2021yyu}) as

\begin{equation}\label{eq:muid}
    \mu_{i,d}=[R_{0,d}(1+f_d\phi_{bkg,d}^{MC}(t_i))+S_m \cos(\omega(t_i-t_0))]M_d\Delta E \Delta t,
\end{equation}

where $M_d$ is the mass of every module, $\phi_{bkg,d}^{MC}$ is the PDF sampled from the MC background evolution in time calculated independently for every module, and $R_{0,d}$ and $f_d$ are considered free parameters. Then, assuming $S_m$=0, for each ten-day time bin $i$ and detector $d$, a Gaussian distribution of mean $\mu_{i,d}$ and standard deviation $\sigma_{i,d}$ is sampled for the $R_{0,d}$ and $f_d$ values derived from the \ANAIS three-year exposure~\cite{Amare:2021yyu}. In this calculation, we correct the constant term (nonvarying rate), $R_{0,d}$, by the percentage of reduction attained with the BDT method (see Table~\ref{tab:bkg}). The mean value $\mu_{i,d}$ in the $i^{th}$ time bin and detector $d$ is computed by evaluating Equation~\ref{eq:muid} at the centre of the time bin $t_i$, whereas the standard deviation $\sigma_{i,d}$ is obtained by multiplying the mean value $\mu_{i,d}$ by the expected relative error for the rate of events at $t_i$. The relative error in the rate of events has been estimated assuming that the live time distribution of \ANAIS is homogeneous over time, as well as the detection efficiency in [1,6] and [2,6]~keV, and considering the background evolution modelled by $\phi_{bkg,d}^{MC}$. In this way, the standard deviation of the modulation amplitude in the $i^{th}$ time bin is estimated by fitting the simulated rates of events up to such time bin to Equation~\ref{eq:muid}.

\begin{figure}
  	\begin{subfigure}[b]{0.5\textwidth}
      \centering
  	  \includegraphics[width=\textwidth]{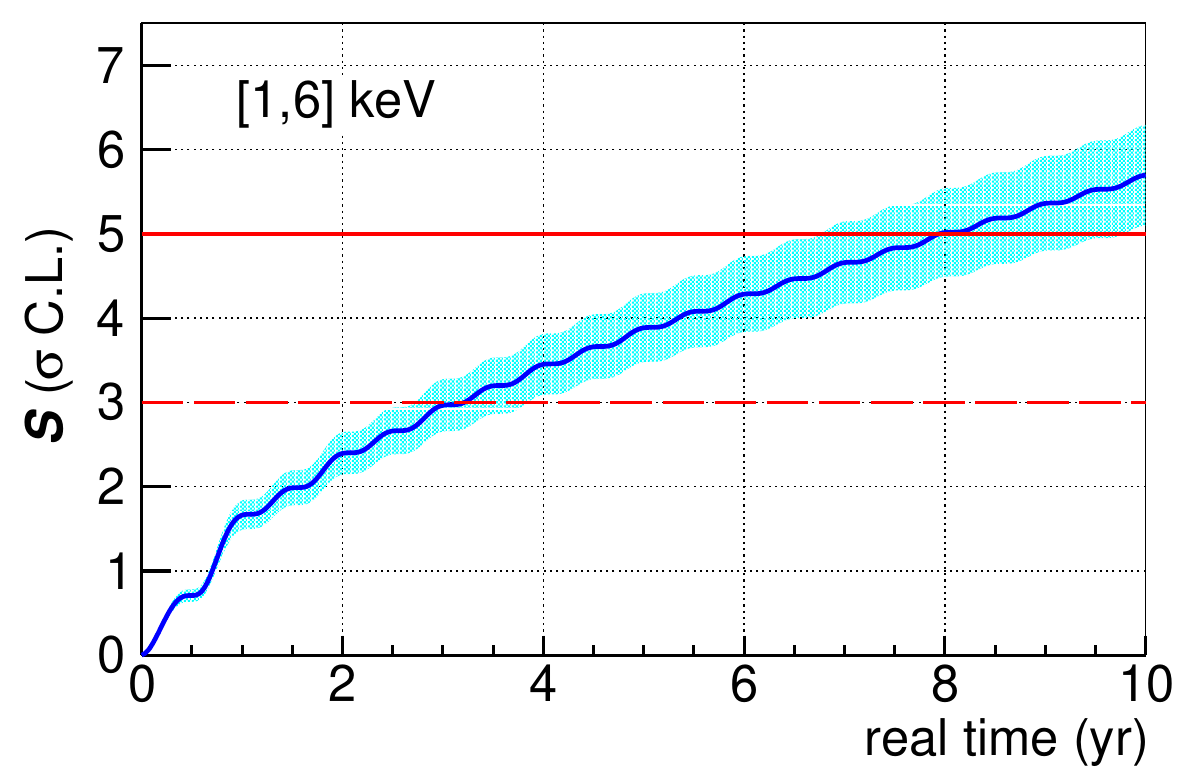}
  	  \caption{\label{fig:sensBDT16}}
  	\end{subfigure}%
  	~ 
  	\begin{subfigure}[b]{0.5\textwidth}
      \centering
  	  \includegraphics[width=\textwidth]{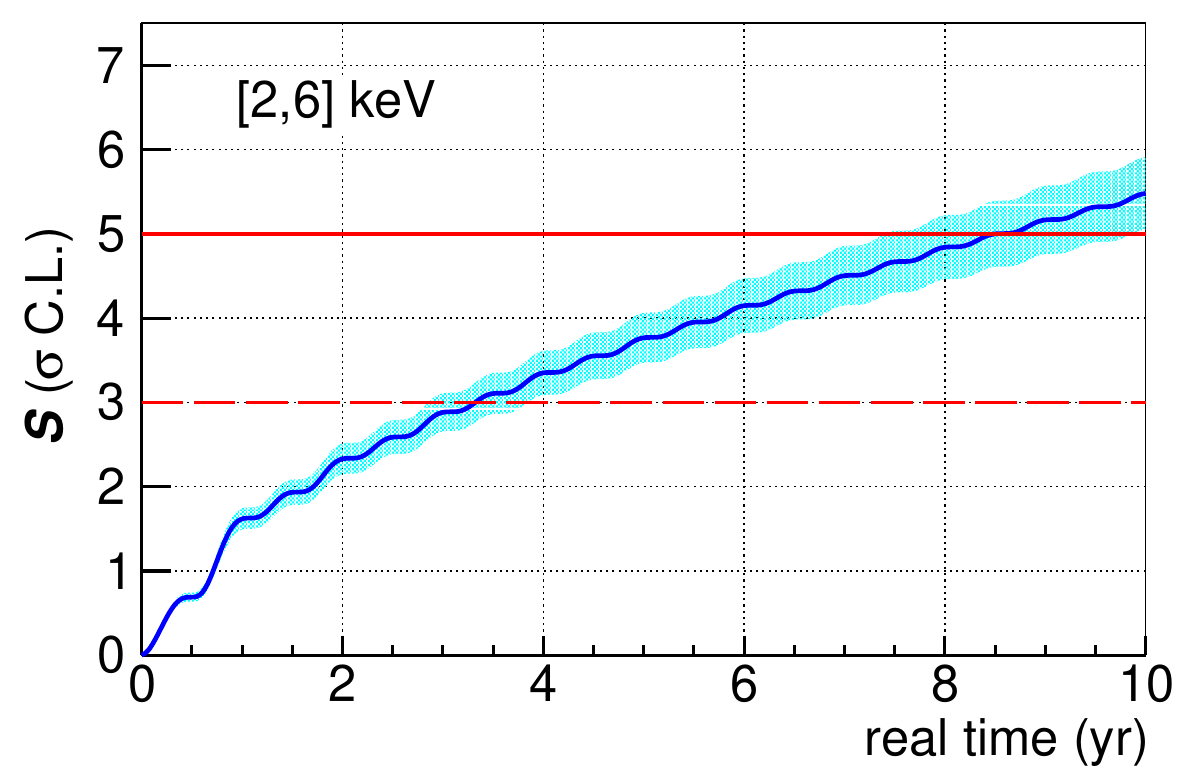}
  	  \caption{\label{fig:sensBDT26}}
  	\end{subfigure}%
   \caption{\ANAIS sensitivity to the \DL signal in $\sigma$ C.L. units as a function of real time in the [1,6] keV (a) and [2,6] keV (b) energy regions considering decreasing background after applying the BDT event selection. The cyan bands represent the 68\% C.L. \DL uncertainty. \label{fig:sensBDT}}
\end{figure}

\begin{figure}
  	\begin{subfigure}[b]{0.5\textwidth}
      \centering
  	  \includegraphics[width=\textwidth]{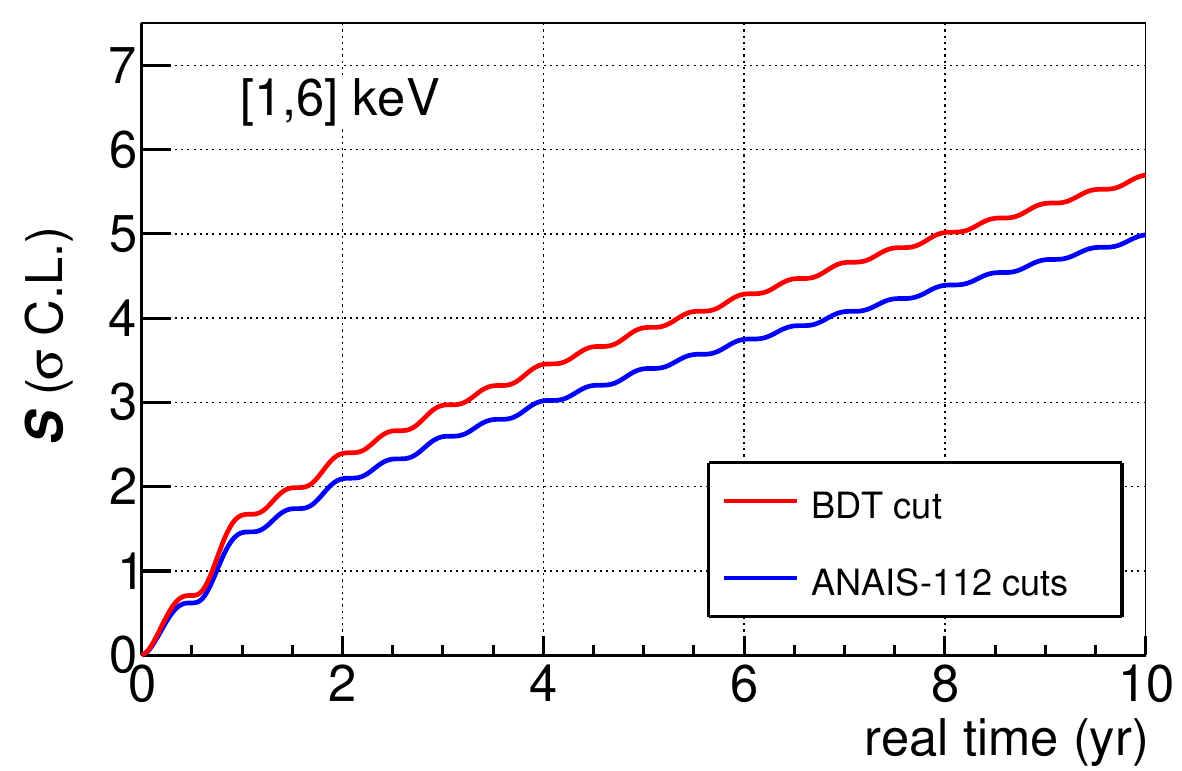}
  	  \caption{\label{fig:sensANAISvsBDT16}}
  	\end{subfigure}%
  	~ 
  	\begin{subfigure}[b]{0.5\textwidth}
      \centering
  	  \includegraphics[width=\textwidth]{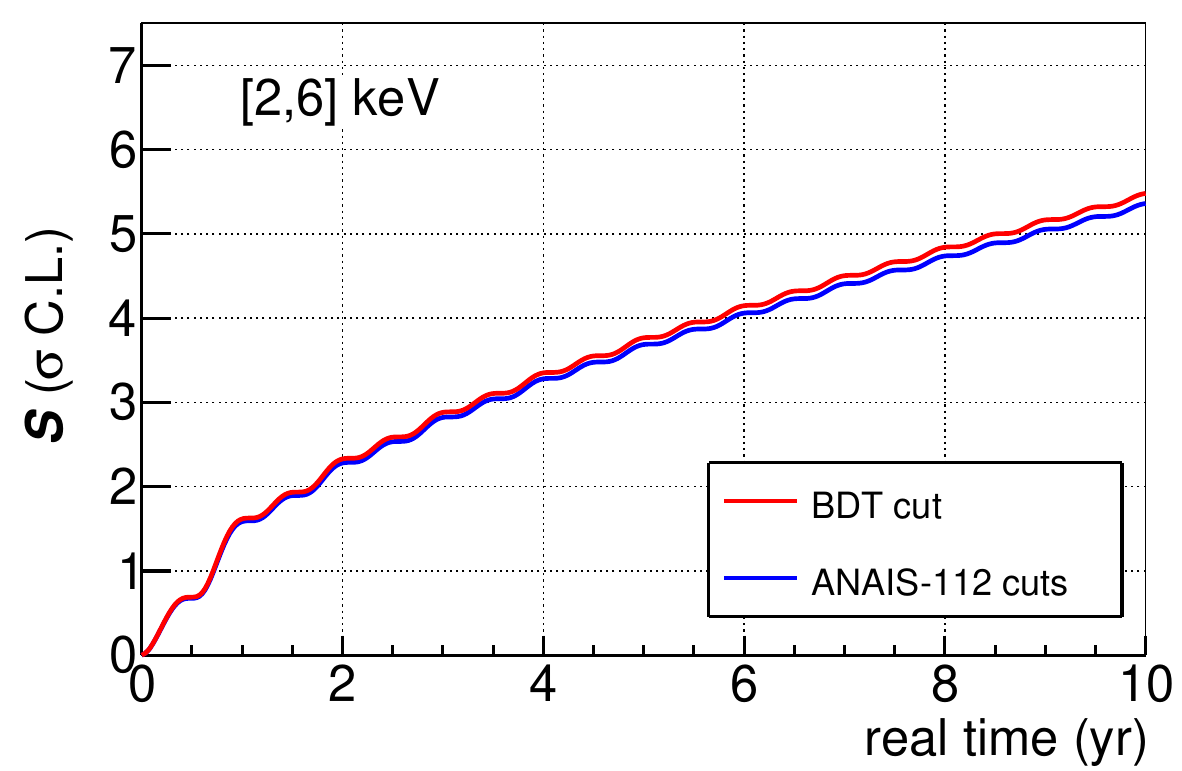}
  	  \caption{\label{fig:sensANAISvsBDT26}}
  	\end{subfigure}%
   \caption{\ANAIS sensitivity to the \DL signal in $\sigma$ C.L. units as a function of real time in the [1,6] keV (a) and [2,6] keV (b) energy regions considering decreasing background. The blue lines show our sensitivity projection derived from applying the previous \ANAIS filtering procedure, whereas the red lines display the expected sensitivity from applying the BDT method, as shown in Figure~\ref{fig:sensBDT}. \label{fig:sensANAISvsBDT}}
\end{figure}

Figure~\ref{fig:sensBDT} (dark blue lines) shows our sensitivity projection for the two studied energy ranges, whereas the cyan bands represent the 68\% uncertainty in $S_m^{\text{DAMA}}$ as reported in~\cite{Bernabei:2018jrt}. Figure~\ref{fig:sensANAISvsBDT} shows the improvement of the new sensitivity estimates using the BDT method with respect to the previous ones~\cite{Amare:2021yyu}. According to our new sensitivity estimates, the \ANAIS experiment would test the \DL annual modulation result around 3$\sigma$ sensitivity by applying this new BDT filtering to the three-year exposure, being possible to achieve 5$\sigma$ sensitivity by extending the data taking for 3--4 more years than the scheduled 5~years which were due in August 2022.
%
%

\section{Conclusions}
\label{sec:conclusions}

A new low-energy noise event filtering protocol based on the boosted decision tree technique has been developed for the \ANAIS experiment. For the BDT training, we have used genuine nuclear recoils from dedicated onsite neutron calibrations as signal events and events coming from a blank module (i.e., identical to the \ANAIS ones, but without scintillating NaI(Tl) crystal) as noise sample, without requiring to use background events at all.

With this filtering procedure, a background level reduction of around 20\% has been achieved between 1 and 2~keV. However, the excess of events below 2 keV with respect to our background model is still present, pointing at a background contribution in this energy range unaccounted for in the model and supporting the convenience of the revision of the \ANAIS background model. 

Moreover, this new filtering increases significantly the efficiency of the event selection, pushing the \ANAIS sensitivity to test the \DL annual modulation result around 3$\sigma$ with three-year exposure. A reanalysis of annual modulation in the first 3 years of \ANAIS\cite{Amare:2021yyu} with the new filtering technique is ongoing and will appear soon. According to our sensitivity projections, \ANAIS is able to reach 5$\sigma$ sensitivity to \DL signal by extending the data taking for a few more years than the scheduled 5~years, which were due in August 2022.


\acknowledgments

This work has been financially supported by MCIN/AEI/10.13039/501100011033 under grant PID2019-104374GB-I00, the Consolider-Ingenio 2010 Programme under grants MultiDark CSD2009-00064 and CPAN CSD2007-00042, the LSC Consortium, and the Gobierno de Arag\'on and the European Social Fund (Group in Nuclear and Astroparticle Physics). Authors would like to acknowledge the use of Servicio General de Apoyo a la Investigaci\'on-SAI, Universidad de Zaragoza and technical support from LSC and GIFNA staff.

\bibliographystyle{JHEP}
\bibliography{ANAIS_BDT}

\end{document}